\documentclass[twocolumn,tighten,astrosymb]{aastex631}

\usepackage{newtxtext,newtxmath}

\usepackage[T1]{fontenc}
\usepackage{ae,aecompl}

\usepackage{graphicx}	
\usepackage{amsmath}	
\usepackage{amssymb}	

\def\gtorder{\mathrel{\raise.3ex\hbox{$>$}\mkern-14mu
    \lower0.6ex\hbox{$\sim$}}}
\def\ltorder{\mathrel{\raise.3ex\hbox{$<$}\mkern-14mu
    \lower0.6ex\hbox{$\sim$}}}

\begin{document}


\title{Direct Collapse to Precursors of Supermassive Black Hole Seeds:\\ Radiation-feedback-generated Outflows}
\shorttitle{Direct collapse to Precursors of SMBH seeds: Radiation Feedback}


\author[0000-0002-2243-2790]{Yang~Luo}
\affiliation{Department of Astronomy, Yunnan University, Kunming, Yunnan 650091, China}

\author[0000-0002-1233-445X]{Isaac~Shlosman}
\affiliation{Department of Physics \& Astronomy, University of Kentucky, Lexington, KY 40506-0055, USA} 
\affiliation{Theoretical Astrophysics, Department of Earth \& Space Science, Osaka University, 1-1 Machikaneyama, Toyonaka, Osaka 560-0043, Japan}

\author[0000-0001-7457-8487]{Kentaro~Nagamine}
\affiliation{Theoretical Astrophysics, Department of Earth \& Space Science, Osaka University, 1-1 Machikaneyama, Toyonaka, Osaka 560-0043, Japan} 
\affiliation{Department of Physics \& Astronomy, University of Nevada, Las Vegas, NV 89154-4002, USA} 
\affiliation{Kavli-IPMU (WPI), University of Tokyo, 5-1-5 Kashiwanoha, Kashiwa, Chiba, 277-8583, Japan} 

\email{luoyang@ynu.edu.cn (YL)}
\email{isaac.shlosman@uky.edu (IS)}

\shortauthors{Y. Luo et al.}

\date{Accepted XXX. Received YYY; in original form ZZZ}



\label{firstpage}

\begin{abstract}
We use high-resolution zoom-in cosmological simulations to model outflow triggered by radiation and thermal drivers around the central mass accumulation during direct collapse within the dark matter (DM) halo. The maximal resolution is $1.3\times 10^{-5}$\,pc, and no restrictions are put on the geometry of the inflow/outflow. The central mass is considered {\it prior} to the formation of the supermassive black hole seed at a redshift of $z\sim 15.9$, and can constitute either a supermassive star (SMS) of $\sim 10^5\,M_\odot$ surrounded by a growing accretion disk or a self-gravitating disk. The radiation transfer is modeled using the ray-tracing algorithm. Due to the high accretion rate of $\sim 1\,M_\odot\,{\rm yr^{-1}}$ determined by the DM halo, accretion is mildly supercritical, resulting in mildly super-critical luminosity which has only a limited effect on the accretion rate, with the duty cycle of $\sim 0.9$. We observe a fast development of hot cavities, which quickly extend into polar funnels and expand dense shells. Within the funnels, fast winds, $\sim 10^3\,{\rm km\,s^{-1}}$, are mass-loaded by the accreting gas. We follow the expanding shells to $\sim 1$\,pc, when the shell velocity remains substantially, $\sim 5$ times, above the escape speed. The ionization cones formed by the central UV/X-ray completely ionize the cavities. Extrapolating the outflow properties shows that the halo material outside the shell will have difficulty stopping it. We therefore conclude that the expanding wind-driven shell will break out of the central parsec and will reach the halo virial radius. Finally, the anisotropic accretion flow on sub-parsec scales will attenuate the UV/soft X-rays on the H$_2$. Hence, the formation of funnels and powerful outflows around, e.g., SMS, can have interesting observational corollaries. 
\end{abstract}




\section{Introduction}
\label{sec:intro}

The discovery of supermassive black holes (SMBHs), with masses of $\sim 10^9\,\rm M_{\odot}$, at $z\gtorder 7$, brought up the question of how these objects could form in less than 750\,Myr from the Big Bang \citep[e.g.,][]{fan03,willott10,mortlock11,wu15,venemans17,banados18,bosman23,larson23}. Moreover, recent detection of an SMBH with the mass of $\sim 4\times10^7\,\rm M_\odot$ at $z \sim 10.3$, has reduced the formation time even further\citep{bogdan23}.

The direct collapse scenario, which involves the baryonic collapse within the dark matter (DM) haloes to form SMBH seeds of $\sim 10^4 - 10^6\,\rm M_\odot$, has been proposed as a viable way to form these SMBHs \citep[e.g.,][]{rees84,haehnelt93,loeb94,bromm03,koushiappas04,begelman06}. An alternative model, that SMBHs grow from  stellar remnant BHs suffers from the problem of the gas feeding, and requires prolonged time intervals of a supercritical accretion\footnote{By supercritical accretion we mean an accretion rate in excess of the Eddington rate.} \citep[e.g.,][]{choi13,madau14,lupi16,li19}. Whereas the model to form the SMBHs by a runaway collapse of compact stellar clusters also has numerous difficulties, namely, the formation of such clusters at high redshifts, the unavoidable stellar slingshot effects, the fine-tuning of the gas feeding, etc. \citep[e.g.,][]{begelman78,devecchi09,lupi14,li17,kroupa20}. 

During the gravitational collapse, the gas accumulates at the center and is expected to form an SMBH seed, either via  the intermediate stage of a supermassive star (SMS; \citep{baumgarte99,shapiro02,shapiro04,begelman06,hosokawa09,begelman10,hosokawa13,woods21}), or bypassing it \citep{begelman09,choi13}. In the former option, the core collapse of the SMS would leave the SMBH seed of $\sim 10-10^5\,M_\odot$, which will grow rapidly via a supercritical accretion \citep{volonteri05,wang06,inayoshi14b,sakurai16b,inayoshi16,takeo20}. In the latter case, the collapse leads to the formation of a rapidly rotating self-gravitating disk \citep{begelman09,choi13, choi15}. The final stages of the direct collapse involve trapping of radiation, and require radiation transfer performed on-the-fly \citep{luo18,ardaneh18}. Therefore, conditions associated with this collapse can generate powerful winds. While various types of outflows have been modeled and observed for the SMBHs in active galactic nuclei (AGN), as well as from massive stars, winds from the precursors of the SMBHs at high redshifts have not been addressed so far. 

In this work, we focus on radiation and thermally-driven outflows from the precursors of the SMBH within direct collapse scenario, i.e., under conditions when the SMBH seed has not yet formed.  

The direct collapse scenario requires a gas of a pristine composition within DM haloes having the virial temperature above the atomic cooling floor $\sim 10^4$\,K \citep[e.g.,][]{bromm03,wise08,regan09,greif11,choi13,latif13,choi15,shlosman16,latif16,inayoshi20}. It also requires that the molecular hydrogen formation is suppressed during the collapse by the presence of a strong ultraviolet (UV) background from nearby star-forming sites, to avoid fragmentation \citep[e.g.,][]{dijkstra08,agarwal12,dijkstra14,yue14,inayoshi15,yue17,maio19, luo20}.  

The accretion rate during direct collapse is governed by the DM halo, $\dot M\sim v_{\rm ff}^3/G$, where $v_{\rm ff}$ is the free-fall velocity in the halo. It can greatly exceed the Eddington limit onto the forming central object \citep[e.g.,][]{begelman06,begelman12,inayoshi20}. The DM haloes with a required virial temperature exceeding that of the cooling floor in atomic gas have a typical mass of $M_{\rm h}\sim 10^7-10^8\,{\rm M_\odot}$, virial radius of $R_{\rm h}\sim 0.1-1$\,kpc, free-fall velocity of $v_{\rm ff}\sim 10-20\,{\rm km\,s^{-1}}$, and mass accretion rate of $\dot M\sim 0.1-1\,{\rm M_\odot\,yr^{-1}}$ \citep[e.g.,][]{begelman09,choi15,luo16,luo18,ardaneh18}. 

In the vicinity of the growing central object, e.g., SMS or self-gravitating accretion disk, where its gravitational potential dominates over that of the DM halo, the Eddington luminosity and accretion rate, $L_{\rm Edd}\sim \dot M_{\rm Edd} v_{\rm ff}^2$, are related by $v_{\rm ff}^2\sim G M_{\rm SMS}/R_{\rm SMS}$, so
\begin{equation}
\dot{M}_{\rm Edd} \sim 0.6\eta^{-1}\beta^{-1} \bigg(\frac{R}{R_{\rm SMS}}\bigg)\, {\rm M_\odot\,yr^{-1}},
\label{Medd}
\end{equation}
where $R_{\rm SMS}\sim 1.3\times 10^{-5}$\,pc is the central accumulation radius for $M_{\rm SMS} = 10^5\,M_\odot$ used in this work, $\beta$ is the cross-section of the radiation--accreting matter interaction in units of the Thomson cross section, which depends on the energy band of radiation, and $\eta$  is the radiative efficiency which is typically taken as $\sim 0.1$ when the SMBH is present. However, before the horizon is established, as in the case of interest here,  $\eta\sim 1$, because all the radiation can escape in principle, and not be absorbed by the horizon. In other words, we define the efficiency of conversion with respect to the kinetic energy of accretion.

Accretion onto the central object means that a growing accretion disk will form, as some angular momentum will be present in the accretion flow. Gravitational torques become inefficient in the vicinity of the central massive object \citep{shlosman16}. Previous studies have shown that super-Eddington accretion could be realized when the photon diffusion time scale is shorter than the advection time scale, and radiation is trapped in the optically-thick accretion flow \citep{begelman78,abramowicz88,jiao11,sadowski11,jiang14,kitaki18,feng19}. Observations also hint at the appearance of a supercritical accretion in AGN \citep{wang14,du15,du16,cackett20,tortosa21}. The released luminosity can reach the Eddington luminosity, and the radiative feedback will have a strong effect on the gas feeding and mass supply \citep[e.g.,][]{wang99,ohsuga05,sadowski16,inayoshi16,kitaki18}. 

Numerical simulations have shown that gravitational collapse within the DM haloes can be divided into two phases: the optically-thin outer region extending down to $\sim 10^{-4}$\,pc, and the innermost optically-thick phase which requires the radiation transfer on-the-fly. To circumvent this,  the inner accretion flow is usually approximated by the adiabatic equation of state \citep[e.g.,][]{becerra15,becerra18}. However, the validity of this approximation was never justified. The 3-D radiation transfer has been only recently implemented and has shown that the evolution pattern differs in this case from the adiabatic one \citep{luo18,ardaneh18}. Note also the 2-D radiation transfer implemented by \citet{sugimura18}, who modeled the subsequent formation of the disk and its evolution using a subgrid model.

The growth of the central mass accumulation has been followed self-consistently only to $\sim 100\,M_\odot$ \citep{luo18,ardaneh18}.  It has revealed the formation of the central object which is not in the hydrostatic equilibrium. The kinetic energy of accretion flow has been deposited below its photosphere, dissipated there and formed hot bubbles which tend to discharge above the photosphere, driving strong outflows. The driving forces behind these outflows have been both thermal pressure gradients and radiation force. Below the photosphere, strong turbulence has developed. Moreover, the comparison between the adiabatic inflow and that with the radiation transfer displays a diverging evolution \citep{luo18,ardaneh18}. 

The accretion rate in these simulations has exhibited a clear supercritical behavior above the photosphere, $\sim 0.1-1\,{\rm M_\odot\,yr^{-1}}$, but the emerging luminosity has been limited to the Eddington luminosity, and varied in the range of $\sim (10^{-3}-1)\,L_{\rm Edd}$, for Thomson cross section.

The late stages of direct collapse, when a large central mass has been accumulated, are computationally consuming. Formation of the SMS or the self-gravitating disk in excess of a few$\times 10^2\,M_\odot$ has never been achieved self-consistently. Typically, the central mass has been assumed to form via 1-D or 2-D simulations, and its evolution is followed by low-dimensional stellar evolution codes. For example, \citet{sakurai16b} used a spherical accretion from a surrounding cloud onto the sink particle representing the SMS.  {\it Ad hoc} 5\% of the gas has been allowed to form an accretion disk around the sink particle, and the structure of this disk has been imposed, which determined the follow-up evolution of this disk. Having a large surface density in the geometrically-thin disk forced it to fragment. Because the authors pointed out that the luminosity of the SMS provides negligible radiation feedback, the accretion luminosity has been ignored. This evolution disagrees with that of fully 3-D simulations of direct collapse, both for isolated cases and in a full cosmological context, when no fragmentation has been observed, even in the absence of radiation feedback \citep{choi13,choi15}, as well as in the context of an accretion disk around a massive sink particle \citep{shlosman16}.

The 1-D accretion to form an SMS in the presence of radiation feedback and accounting for the molecular hydrogen has been performed by \citet{sakurai20}. However, the UV feedback has dissociated the H$_2$ at small radii, and its effect was found to be only transient, by temporarily increasing the gas temperature and thermal pressure.

Accretion onto the BH differs from accretion onto a less compact object analyzed here by the emerging spectrum of accretion luminosity --- a power law versus blackbody, and the presence of the hard X-ray radiation in the former case. This results in a substantial difference in the interaction of radiation with the accreting gas. 

While we do not deal with accretion onto the SMBHs here, a clear analogy exists between accretion and feedback onto the SMBHs and on the massive objects which grow in the central regions of DM halos. We review them briefly below. 

Radiation feedback has been used in previous works modeling gravitational collapse, but with such a low spatial resolution that associated physical conditions had little relation to those encountered in direct collapse, or dealt with accretion onto the SMBH. For example, \citet{johnson11} has barely resolved the Bondi accretion radius onto the BH of $\sim 10^5\,M_\odot$, but did not resolve by far the region of the collapse where the angular momentum becomes important and may lead to the formation of an accretion disk. \citet{latif18} have modeled radiation feedback from a BH of $\sim 10^5\,M_\odot$ inserted at $z=7.5$, but used the maximal spatial resolution of 3.6\,pc, and hence suffered from the same problem. \citet{milosavljevic09} have implemented the ionizing radiation feedback but ignored the angular momentum, thus modeling a purely radial accretion flow. \citet{park12} analyzing the radiation feedback onto spherical accretion finds no significant effect of the angular momentum on the flow. Finally, \citet{park17} have analyzed the role of turbulence in the accretion flow and concluded that its effect is not significant.

Numerical modeling concerning the accretion onto the SMBHs associated with radiation feedback has shown that the surrounding gas gets photo-evaporated, and the accretion rate is strongly suppressed \citep{pelupessy07,sijacki09,booth09}. Works on the effect of radiation feedback from the SMBH seeds found that the accretion rate drops to $\sim 10^{-5}\,\rm M_\odot yr^{-1}$, which is below the Eddington accretion rate for the central SMBH with a mass of $10^5\,\rm M_\odot$ \citep{johnson11,chon18,latif18,regan19}. The possibility of maintaining hyper-critical flow in a 1-D spherical accretion has been modeled as well \citep{inayoshi16}.  

\citet{smidt18} has modeled accretion onto SMBH in a large computational box of $100 h^{-1}$\,Mpc on the side, and three nested grids of $25 h^{-1}$\,Mpc have been applied. Direct collapse within a DM halo at $z\sim 19$ has been followed onto the sink particle of $10^5\,M_\odot$ representing a BH with the entire radiation feedback produced at 1\,keV. However, the relevant physics in the central region was completely unresolved, with the maximal resolution of 35\,pc (comoving). Moreover, the radiation feedback, the accretion disk and the associated outflow, have been chosen as a subgrid model of an $\alpha$-disk.

\citet{regan19} have modeled accretion onto the SMBH seed of $\sim 10^5\,M_\odot$ with a highest spatial resolution of $\sim 2.5\times 10^{-4}$\,pc (which is about a factor of 20 lower than used in this work, see section\,\ref{sec:numerics}). The radiation feedback from the SMBH accreting at the super-Eddington rate has been modeled using the ray-tracing algorithm, but the ionizing radiation has been ignored, and the simulation focused entirely on the mechanical feedback from the collimated jet. The jet and its effects on the accretion have been not detectable at $\gtorder 0.1-1$\,pc, but the SMBH growth rate has been severely suppressed. 

In the present work, we have avoided including subgrid physics. The outflow from the central object modeled here represents a wind, partially collimated and driven by the resolved forces --- radiation and thermal pressure gradients.    

In this work, we utilize the zoom-in 3-D cosmological radiative hydrodynamics simulations and study the evolution of a central mass accumulation in the presence of radiation feedback, including the photoionization input from the UV and X-ray photons. This central mass accumulation can represent an SMS or self-gravitating disk, with the resolution of $\sim 1.3\times 10^{-5}$\,pc. We employ a ray-tracing algorithm to model the propagation of the ionizing photons on the pristine gas. The effects of radiation pressure on this gas are also included.

This paper is structured as follows. Section\,2 describes the numerical methods used here, the initial cosmological conditions, and methods to calculate accretion rate and luminosity. Our results are presented in Section\,3. Finally, we discuss and summarize this work in Sections\,4  and 5.

\section{Numerical Method}
\label{sec:numerics}

For simulations of direct collapse within DM haloes, and the modeling of the radiation feedback from a central sink particle, we perform 3D cosmological simulations using the Eulerian adaptive mesh refinement (AMR) code Enzo-2.5 \citep{bryan97,norman99,bryan14}. The gravitational dynamics is calculated by a particle-mesh $N$-body method \citep{colella84,bryan95}. The piece-wise parabolic method, which is an improved form of the Godunov method \citep{colella84}, is implemented to solve the hydrodynamics equations. It makes use of the multi-grid Poisson solver for self-gravity calculations. More details of our simulations could also be found in \citet{luo16}, \citet{luo18}, and \citet{ardaneh18}.

\subsection{Simulation Setup}
\label{sec:setup}

In the first step, we start our lower resolution, pathfinder simulation at redshift $z = 199$ with cosmological initial conditions of $1\,h^{-1}$ Mpc comoving box and the root grid  dimension of $128^3$, generated by the MUSIC package \citep{hahn11}, and run it without the AMR and baryonic component until $z = 10$. Then we select an appropriate DM halo by using the HOP group finder \citep{eisenstein98}. Next, we generate a zoom-in DM halo with $1024^3$ effective resolution in DM and gas, centered on the selected halo position. 

We choose the zoom-in region to be large enough to avoid contamination with the high-mass, low-resolution DM particles. Within the zoom-in region, a number of $\sim 10^7$ refined DM particles is used, which yield an effective DM particle mass resolution of about 99\,M$_{\odot}$. The baryon resolution is set by the size of the grid cells. 

The grid cells are adaptively refined based on the following three criteria: baryon mass, DM mass, and Jeans length. A region of the simulation grid is refined by a factor of $2$ in length scale, if the gas or DM densities become greater than $\rho_0 2^{(3+\alpha) l}$, where $\rho_0$ is the density above which the refinement occurs, and is obtained by assuming the density of the cell exceeds 4 times the mean density of the subgrid. Here $l$ is the refinement level and $\alpha$ is set to $-0.2$, which makes the refinement super-Lagrangian \citep{bryan14}.

We impose the condition of at least 24 cells per Jeans length in our simulations, so that no artificial fragmentation would take place \citep{truelove97}. In all simulations, we set the maximum refinement level to $25$, which provides about $1.3\times10^{-5}$\,pc physical resolution when the maximum refinement level is reached. 
 
The gas chemistry is assumed to be  of the primordial composition, and we use the publicly available package GRACKLE-3.1.1\footnote{https://grackle.readthedocs.org/} \citep{bryan14, smith17} to follow thermal and chemical evolution of the collapsing gas. GRACKLE is an open-source chemistry and radiative cooling/heating library suitable for use in numerical astrophysical simulations. 

We use Planck\,2015 for cosmology parameters \citep{planck-collaboration16}: $\Omega_{\rm m}$ = 0.3089, $\Omega_\Lambda$ = 0.6911, $\Omega_{\rm b}$ = 0.04859 $\sigma_8$ = 0.8159, $n_{\rm s}$ = 0.9667, and the Hubble constant $h$ = 0.6774 in units of $100\,\rm km\,s^{-1}\,Mpc^{-1}$.

We also use $R$ to abbreviate the spherical radii, and $r$ for cylindrical radii.

\subsection{Sink Particle Method}
\label{sec:sink}

The long-term simulation of the SMS evolution with radiation transfer is very computationally expensive. Therefore, an algorithm of sink particles is sometimes introduced. The central mass accumulation is represented by the sink particle in cells where the maximum refinement level is reached \citep{krumholz04,federrath10,hubber13,shlosman16}. In this work, we treat the central mass accumulation as the SMS, the precursor of the SMBH seed. The introduction of sink particles at the highest resolution helps to decrease the dynamic range and hence to reduce the computational time and extend the simulation time to a longer time interval. In this work, we largely follow the prescription of \citet{shlosman16}.

The sink particle is allowed to form when a particular cell violates the refinement criterion, and this happens at the highest refinement level in the collapsing flow. As the gas density increases and then the maximum refinement level is reached, the sink particle is inserted at the center of the cell if the following criteria are met: (i) The cell is at the highest refinement level, (ii) The cell exceeds the Jeans density, (iii) The flow around the cell is converging along all axes, (iv) The cooling time of the cell is less than the free-fall time \citep{federrath10, regan18}. Its initial mass is computed from the mass exceeding the maximum allowed density at a maximum refinement level. Under this method, each cell density has a maximal value that does not violate the refinement criterion. The initial velocity of the sink particle is determined from the linear momentum conservation. After formation, the sink accretes the gas from its host cell according to the following prescription \citep{ruffert94,krumholz04},
\begin{equation}
\dot M_{\rm sink} = 4\pi \rho_{\rm out}\, R_{\rm B}^2 \sqrt{1.2544 c_{\rm s}^2 + v_{\rm rel}^2},
\end{equation}
where $\dot M_{\rm sink}$ is the mass growth rate of the sink, $c_{\rm s}$ is the sound speed in the parent cell, $v_{\rm rel}$ is the relative velocity between the host cell and the sink particle, and $R_{\rm B} = GM_{\rm sink}/(c_{\rm s}^2 + v_{\rm rel}^2)$ is the Bondi accretion radius. Finally, $\rho_{\rm out} = \rho_{\rm cell}\,{\rm min}\,[1.0, (l_{\rm cell}/R_{\rm B})^{1.5}]$. We find that the relative velocity for massive sink particles is $(v_{\rm rel}/c_{\rm s})^2 \ll 1$ and can be ignored.

Under these conditions, accretion onto the parent cell hosting the sink particle is not isotropic. In other words, we resolve the Bondi accretion radius. This is very important as one can encounter situations when accretion and outflow occur simultaneously along different directions. The typical situation is when the outflow is directed at a solid angle along the rotation axis of the sink, while the inflow comes from near the equatorial plane of the neighboring accretion disk.

The sink particles are also allowed to merge in order to accurately estimate their mass growth rate and to reduce the computational cost. Typically, sink particles can be interpreted as overdense gas clumps whose internal evolution we ignore. The condition for two sink particles to merge is fulfilled when their separation becomes smaller than $4l_{\rm cell}$. The less massive sink particle is assumed to merge with the more massive one. This merging process is insensitive to the definition of the critical separation between particles. 

In this paper, we skip the discussion of the cosmological phase of evolution as it is similar to our previous simulations \citep[e.g.,][]{shlosman16,ardaneh18}, and only mention the essential differences coming from the increased resolution. Here, the direct collapse was triggered around $z\sim 15.9$, when the virial mass and radius of the parent DM halo reached $M_{\rm h} \sim 2.4\times10^{7}\,M_\odot$ and $R_{\rm vir}\sim 0.6$\,kpc. Sink particles of $\sim 10\,M_\odot$ have been continuously forming, merging, accreting the surrounding gas, and falling to the center, where the fastest growing sink has been found. The relative velocity of this central sink with respect to the computational box has been quickly decaying, and after it reached a few\,$\times 10^3\,M_\odot$, it became negligible compared to the sound speed in the neighboring cells. 

\subsection{The properties of the central mass accumulation}
\label{sec:mass-size}

 There is no general agreement on the radius and internal structure of an SMS, including the position of its photosphere, especially when accretion and spin are involved. Even the existence of the SMS has not been established so far \citep[e.g.,][and references therein]{hirschi17} --- the SMS can be completely by-passed, and the central object can follow the evolution of a self-gravitating disk \citep[e.g.,][]{begelman09,choi13}.

Note that the formation of the central object and its evolution should be closely related to an important characteristic radius --- the trapping radius for the emerging radiation flux, which depends only on the mass accretion rate, $\dot M_{\rm acc}$, is given as

\begin{equation}
   R_{\rm t} =  \frac{\dot M_{\rm acc}\sigma_{\rm T}}{4\pi m_{\rm p} c} \sim 10^{-5} \left(\frac{\dot M_{\rm acc}}{0.1\,M_\odot\,{\rm yr^{-1}}}\right) 
   \quad {\rm pc} ,
\end{equation}
    where $\sigma_{\rm T}$ and $m_{\rm p}$ are the Thomson cross section and the proton mass. For a central object forming inside the trapping radius, the issue of nuclear burning becomes irrelevant, as the flow automatically generates the Eddington luminosity \citep{begelman78b}.

Modeling the SMS with $10^5\,M_\odot$ as a zero-age main sequence (ZAMS) star, and ignoring the effect of accretion and spin, results in the radius of  $R_{\rm SMS}\sim {\rm few}\,\times 10^{-6}$\,pc \cite[e.g.,][and references therein]{fuller86}. For a fast accreting SMS, with $\sim 0.1\,M_\odot\,{\rm yr^{-1}}$, the radius of the photosphere was calculated at $\sim 6\times 10^{-6}\,(M/10^5\,M_\odot)^{0.5}$\,pc \citep{begelman10}. In the 1D simulations imposing a spherical symmetry and hydrostatic equilibrium, as well as including the contribution of the H$^-$ opacity, this radius was claimed to be as large as $1.8\times10^{-3}$\,pc \citep[e.g.,][]{hosokawa13,haemmerle19,herrington23}. For these models, the effective photospheric temperature of a bloated star was found to be less than $10^4$\,K. Therefore, its ionizing luminosity can hardly have an effect on its environment. More importantly, these simulations ignore the effect of an angular momentum present in the accreting matter, and are unable to follow the evolution of an accretion disk forming around the SMS.

The 3-D simulations with a radiation transfer in the flux-limited diffusion (FLD) approximation, for direct collapse both in isolated DM halos and in the cosmological context, have shown a more complex evolution of the growing SMS. For masses up to $\sim 100\,M_\odot$, the photosphere has been found to lie at $\sim 10^{-5}$\,pc with an effective temperature of $4.4\times 10^4$\,K \citep{luo18,ardaneh18}.  These simulations included the H$^-$ opacity and found it has no effect on the position of the photosphere due to its sensitive dependence on the temperature. Moreover, these objects have been found far from being in the hydrostatic or thermal equilibrium, and their spin has clear dynamical consequences on the internal structure of the object. Hence, it is not clear whether the photospheric radius will experience growth in the mass range of $100\,M_\odot-10^5\,M_\odot$ --- future 3-D simulations would address this issue. 

The 3-D simulation of an isolated collapsing gas cloud with no angular momentum but with induced turbulence, and using the M1 enclosure \citep{rosdahl15}, have shown nearly the same photospheric radius of $2.2\times 10^{-5}$\,pc \citep{kimura23}, compared to $1.3\times 10^{-5}$\,pc of \citet{luo18} and \citet{ardaneh18}, but with a lower photospheric temperature. However, the angular momentum resulting from artificially introducing the turbulent cells differs from the angular momentum developing from cosmological initial conditions. Note that \citet{luo18} compared and found little difference in the radiation hydrodynamical simulations between the M1 enclosure and the FLD algorithms. 

In view of the general uncertainty with the size of the SMS and its structure, we choose the SMS (or the self-gravitating disk size)  as $1.3\times 10^{-5}$\,pc in this work. 

\subsection{Radiation transfer: the adaptive ray method}
\label{sec:rays}

We set a critical mass threshold for the sink particle, $M_{\rm sink}= M_{\rm crit} = 10^5\,M_\odot$, above which we turn on the radiation feedback. This occurs at $z\sim 15.5$, when the parent DM halo has reached a virial mass of $M_{\rm h}\sim 2.5\times 10^7\,M_\odot$. This has been done in order to decrease the computational time. In the following, we define this time as $t = 0$. The duration of the numerical run with the radiation feedback presented here is $3.2\times 10^3$\,yr, which allows us to resolve the number of duty cycles determined by the inflow-outflow cycle. By this time, the outflow has surpassed the distance of 1\,pc from the sink, and we analyze its subsequent fate based on the conservation laws.

Our usage of radiation feedback starting only when the sink mass has reached $10^5\,M_\odot$ is of course an approximation to reduce the computational time. However, it does not change the results of this simulation because, as we show, the steady conditions of the run prevail. Similar effects found in this work are expected to operate when the sink mass is smaller, but the essence will not change.

The radiation feedback is modeled by using the adaptive ray tracing in the MORAY module \citep{wise11}. The radiation field is obtained by integrating the radiation transfer equation along the ray which is considered as monochromatic. 

By using a photon-conserving radiation transfer algorithm, the spatially-adaptive ray tracing method could accurately model the radiation propagation from point sources on a computational grid. The rays are initialized at the sink particle with luminosity equally spread across a number of $N_{\rm pix} = 12\times4^3$ rays. To keep the accuracy of the method, the minimum number of rays is required to be at least $5.1$ rays per cell.

We approximate the central mass concentration by the SMS or by a self-gravitating spinning entity, i.e., disk. The internally generated thermonuclear luminosity (in the case of the SMS) of $M_{\rm sink} = 10^5\,{\rm M_\odot}$ is assumed at the Eddington limit, $L_{\rm nuc}\sim L_{\rm Edd}\sim 1.3\times 10^{43}\,{\rm erg\,s^{-1}}$, and remains close to the accretion luminosity (see below). 

The effective blackbody temperature of $L_{\rm nuc}$ is assumed to peak at $5.8\times 10^4$\,K. For the accretion luminosity, $L_{\rm acc}$ (section\,\ref{sec:feedback}), we use a blackbody spectrum as well with a fixed effective temperature of $6.4\times 10^4$\,K. This corresponds to the average mass accretion rate of $1\,M_\odot\,{\rm yr}^{-1}$. 

The radiation transfer proceeds using three characteristic frequency bins: 13.6\,eV and 20\,eV in the UV, and 0.1\,keV in the soft X-rays for the combined nuclear and accretion luminosity. The inclusion of the X-ray component is important to account for the photoelectric heating in the energy equations and relevant chemistry.
 
Hence, the secondary ionizations \citep{shull85} and Compton heating on electrons from X-rays \citep{ciotti01} are included. The radiation field is coupled to the hydrodynamics solver, and the new ionization fractions and gas energies after every radiative transfer timestep is calculated. The radiation transfer timestep has been assumed as the timestep of the finest AMR level.

The radiation acceleration of the gas as a result of the total radiation acceleration for a cell is calculated as
\begin{equation}
 d{\bf a} = \frac{P}{\rho V_{\rm cell}} \frac{d{\bf p}}{dt}, 
\end{equation}
where $P$ is the photon number  flux along a single ray striking the cell, {\bf p} is the linear momentum in radiation, $dt$ the timestep in the ray-tracing method, $\rho$ is the gas density in a cell, and $V_{\rm cell}$ is the cell volume \citep{wise11}.

The effect of the inverse Compton cooling by the cosmic microwave background is included as well.

\subsection{Radiation feedback from the sink particle}
\label{sec:feedback}

The radiation force can be exerted in principle by the two sources of radiation in our simulations. Firstly, the internally-generated radiation by the nuclear burning, e.g., within the SMS, which is limited by the Eddington luminosity (section\,\ref{sec:rays}). Secondly, the luminosity is generated by the accretion flows onto the sink particle. Below, we describe the way the emerging luminosity is calculated for this case. While the nuclear burning is expected to be steady,  the inflow rate toward the sink particle and the associated luminosity are expected to be bursty.

In direct collapse, the mass inflow rate, $\dot{M}_{\rm acc}\sim v_{\rm ff}^3/G$, is determined on the scale of the virial radius of the parent DM halo, where $v_{\rm ff}$ is the free-fall velocity. On these large scales, the angular momentum in the gas is very low and cannot affect the collapse, but on sub-pc scales, it can become important, depending on the efficiency of the gravitational torques acting on the gas \citep{begelman09,choi13}. On smaller scales, the inflow is separated into quasi-spherical accretion and disk accretion.

The fate of the large-scale infall on sub-pc scales can be two-fold: some of the material is accumulated outside the sink as an accretion disk, while some of the inflow will be converted into the outflow, $\dot M_{\rm out}$, namely,
\begin{equation}
    \dot M_{\rm sink} = \dot M_{\rm acc}  - \dot M_{\rm out}.
\end{equation}
\label{eq:mdot_sink}

In our simulations, the inflow rate $\dot{M}_{\rm acc}$ can exceed the Eddington accretion rate by a factor of a few. Using the sink particle growth rate, $\dot{M}_{\rm sink}$, we calculate the emerging accretion luminosity as originally suggested by \citet{shakura73} and implemented with various small modifications thereafter \citep[e.g.,][]{watarai00,watarai01,mineshige00,du15}: 
\begin{equation}
    L_{\rm acc} = [1+{\rm ln}(\dot{M}_{\rm sink}/\dot{M}_{\rm Edd})] L_{\rm Edd},
\label{luminosity}     
\end{equation}
where the Eddington luminosity is $L_{\rm Edd} \sim 1.3\times10^{43} (M_{\rm sink,5}/M_{\odot})\rm \,erg\, s^{-1}$, where $M_{\rm sink,5} = M_{\rm sink}/10^5{M_\odot}$, and $\dot{M}_{\rm Edd}$ is the Eddington accretion rate (Eq.\,\ref{Medd}).

While the equation\,(\ref{luminosity}) has been designed for the supercritical accretion disks around black holes, clear similarities exist with our case. We adopt it to emphasize that both the accretion rate and the emerging luminosity are only mildly supercritical. Finally, the total luminosity generated by accretion and thermonuclear burning is $L_{\rm tot} = L_{\rm nuc} + L_{\rm acc}$.

During the collapse, the sink particles merge and are accreted by the center where a single sink is growing. After the massive sink at the center exceeds the threshold $M_{\rm crit}$ and the radiation feedback is activated, no additional sink particles are allowed to form. This approximation has been verified as having a negligible impact on the growth of the central sink, and saves computational time \citep{shlosman16}.

\section{Results}
\label{sec:results}

\begin{figure}
\center
\includegraphics[width=0.45\textwidth,height=0.5\textheight,angle=0]{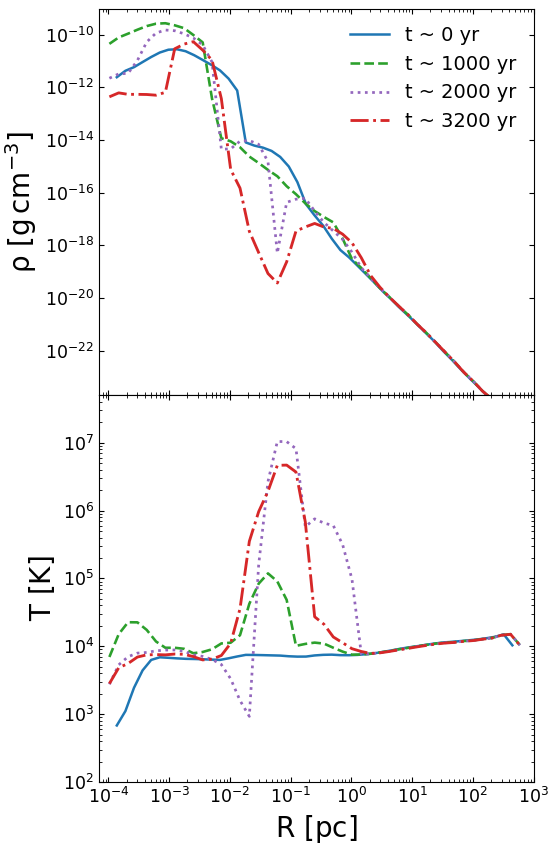}
\caption{Radial profiles of density (top frame) and  density-weighted temperature (bottom frame) within the parent DM halo, calculated in {\it spherical} shells within the DM halo at $t=0$, {\it prior} to triggering the radiation feedback, and at subsequent times with active feedback. The sink particle mass is $\sim 10^5\,M_\odot$. }
\label{fig:initial}
\end{figure}

\begin{figure}
\center
\includegraphics[width=0.45\textwidth,height=0.5\textheight,angle=0]{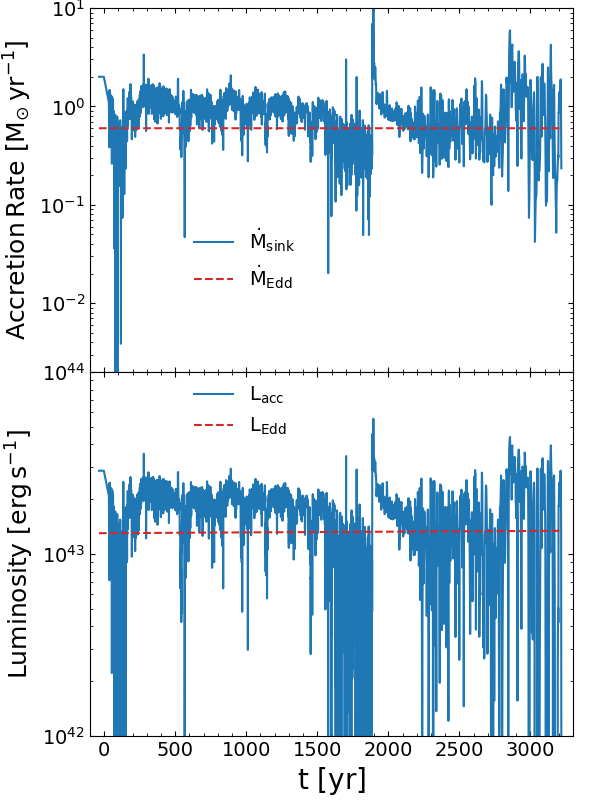}
\caption{Growth rate of the sink particle due to accretion (top frame) and the accretion luminosity from the sink and the surrounding accretion disk (bottom frame). The time $t=0$ corresponds to the time when the sink particle reaches $10^5\,\rm M_\odot$ and when the radiation feedback has been initiated. Prior to this, the sink particle grew from $10\,M_\odot$ to $10^5\,M_\odot$ without feedback. Its growth is negligible over the time of the run with radiation feedback, $3\times 10^3$\,yr. Red-dashed lines show the Eddington accretion rate onto the sink assuming only the electron scattering opacity (top panel), and the Eddington luminosity (bottom panel). The observed noise in the accretion rate is two-fold: the natural variation and the radiation feedback. Both lead to episodic accretion. The outflow generated by the feedback does not totally suppress the accretion, and the sink particle can still grow at a high rate.}
\label{fig:time_rate}
\end{figure}
\begin{figure}
\center
\includegraphics[width=0.46\textwidth,height=0.75\textheight,angle=0]{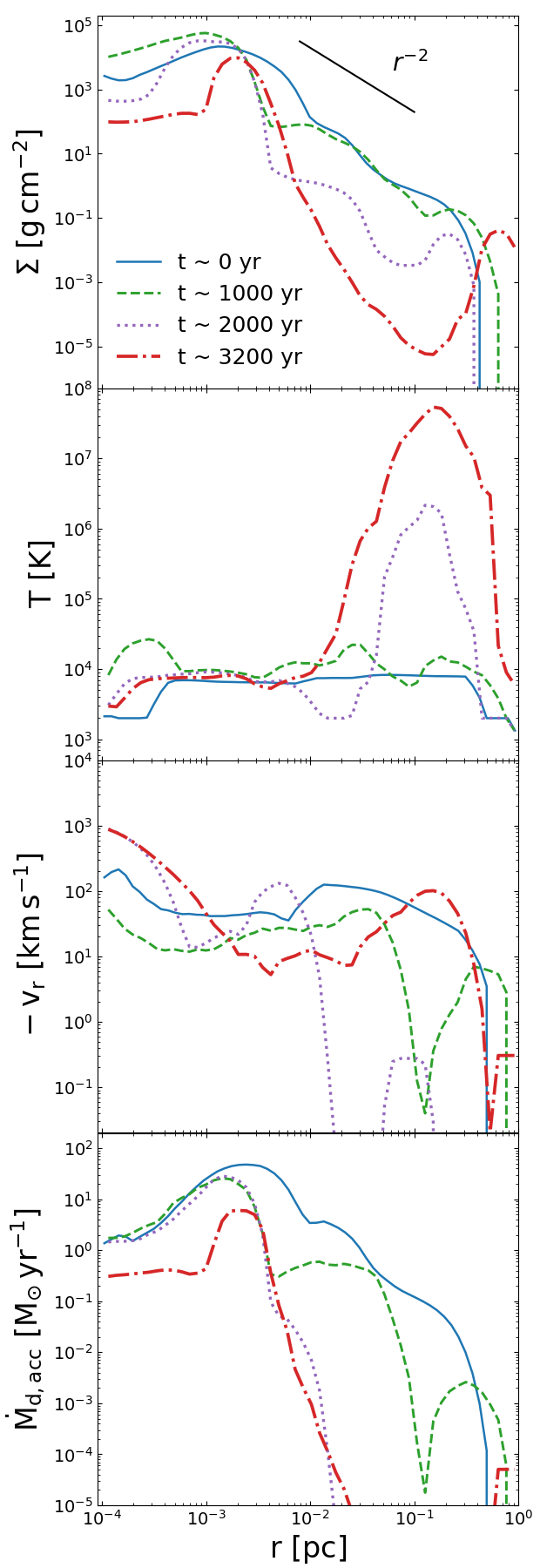}
\caption{Accretion disk structure around the sink particle forming on a scale of $\sim 0.1$\,pc. Shown are the cylindrical radial profiles of the surface density, $\Sigma(r)$ (top panel), and the subsequent frames of the temperature, $T(r)$, inflow velocity, $v_{\rm r}$, and the mass accretion rate in the disk, $\dot M_{\rm d,acc}$, as a function of the cylindrical radius at different times. Note that accretion proceeds also at other polar angles --- this accretion is not included in the disk accretion rate shown here.}
\label{fig:disk}
\end{figure}

We start by analyzing the accretion flow onto the sink and the formation of the accretion disk around it. As a next step, we follow the effects of the radiation feedback and the development of the radiation-driven outflow. Finally, we present the ionization map of the region. Our figures present the time-dependent evolution of various functions during $t\sim 0-3,200$\,yr, and radial profiles of these functions in the range of $r\sim 10^{-4}-1$\,pc and sometimes beyond. While the resolution used here is $1.3\times 10^{-5}$\,pc, the short timescales corresponding to $r\sim 10^{-5}-10^{-4}$\,pc makes the radial profiles there unreliable, and, therefore, we omit them.

\subsection{Accretion onto the sink}
\label{sec:a_sink}

When the central sink particle mass has exceeded $\sim {\rm few}\times 10^3\,M_\odot$, the infalling gas forms a disk around it. We have measured the large-scale accretion flow and confirmed that it is determined by the DM halo, remaining in the range of $\dot{M}_{\rm acc}\sim 1\,M_\odot\,{\rm yr}^{-1}$. On sub-pc scales, we find that the accretion rates differ due to various processes, such as the growth of the sink particle and the formation of an accretion disk around it. The latter one is due to the angular momentum in the accreting matter --- low angular momentum gas is accreted first, while accretion of gas with larger angular momentum is delayed. In this work, we focus on the processes dominating the gas dynamics in the sub-pc region after the sink particle has reached the mass of $M_{\rm sink} = 10^5\,M_\odot$.  

Figure\,\ref{fig:initial} displays the basic parameters of the accretion flow within the DM halo at time $t=0$, i.e., just prior to the radiation feedback has been initiated, and at three subsequent times with the feedback being active. The density and temperature profiles averaged in the {\it spherical} shells are displayed as well. At $t=0$, the density profile reflects the central density core of $R\sim 10^{-2}$\,pc --- which corresponds to the size of the accretion disk, which is followed by a decreasing density with $\sim R^{-2}$, extending to the halo virial radius. 

At later times, the outflow generates dense shells expanding outwards, surpassing 1\,pc by the end of the run, and generating cavities inside the dense shells. The outflows have been triggered by the radiation force and thermal pressure gradients from the vicinity of the sink and the surrounding disk. The formation of the low-density cavities is reflected in a steeper density gradient within few$\times 10^{-2}$\,pc and  expanding dense shell. We focus on the evolution of these shells in the next sections.

The temperature profile appears to be flat at $t=0$, around the cooling floor of $\sim 8,000-10,000$\,K. Only close to the sink particle, it shows a dip initially, which levels off subsequently. At the later times, the temperature shows lower and higher values due to the developing outflow, associated with dense shells and low-density cavities, respectively. The hot gas in these cavities appears to envelop the disk raising the temperature inside. The central parsec has been affected as well by the radiation escaping from the inner region. Averaging the flow in spherical shells can be misleading as it does not allow separate conditions in a very anisotropic inflow and outflow. The alternative way is to follow the inflow and outflow along specific rays.

Figure\,\ref{fig:time_rate} displays the growth rate of the sink due to the accretion (the blue solid line), which experiences rapid variability, occasionally dropping well below the Eddington rate. Around $t\sim 100$\,yr, the accretion rate dips by more than three orders of magnitude, but is quickly restored, with a smaller amplitude variability persisting. 

The average accretion rate remains around $1\,M_\odot\,{\rm yr^{-1}}$, with a quasi-periodic period of $\sim 200$\,yr modulating the rapid variability. This has been verified by the Fourier analysis of the accretion rate until $t\sim 1,500$\,yr. Such characteristic timescale in tandem with the typical inflow velocity in the accretion disk, as we present below, point to a characteristic distance of $\sim 2\times 10^{-2}$\,pc, which is the disk size at that time. Hence the feeding process of the disk is the one affecting the accretion rate during this time period.     

At $t\sim 1,500$\,yr, the accretion rate drops abruptly  below the Eddington rate for a prolonged time of $\sim 300$\,yr. After this time, the accretion rate experiences a spike in growth to about $\sim 10\,M_\odot\,{\rm yr^{-1}}$ and a slower decay back to the Eddington accretion rate of $\sim 0.6\,M_\odot\,{\rm yr^{-1}}$. The variability amplitude increases after this event and is associated with the expanding hot bubble which affects the influx of cold material towards the center, as we detail in the subsequent sections.

The red dashed line in Figure\,\ref{fig:time_rate} shows the corresponding Eddington accretion rate with the dominating electron scattering opacity. The sink particle is growing at a rate of $\sim 0.2-2\, M_\odot\, {\rm yr}^{-1}$ initially, which corresponds to a mildly super-critical accretion rate. The accretion luminosity has been estimated using Equation\,\ref{luminosity}, and is about a factor of 3 above the Eddington luminosity, $L_{\rm acc}\sim 3L_{\rm Edd}\sim 4\times 10^{43}\,{\rm erg\,s^{-1}}$.  

The effect of the radiation feedback can be manifested by a sharp decrease of the accretion rate at different times, even below the Eddington rate, most notably at $t\sim 100$\,yr, $\sim 1,500-1,800$\,yr, $2,200-2,700$\,yr, and around 3,000-3,200\,yr. Note, that in Figure\,\ref{fig:time_rate}, we display only the accretion luminosity and not the total one. 

The accretion rate as a function of time after $t = 0$ also shows a larger variability compared to $t < 0$. Even though the feedback could decrease the accretion substantially, the response of forming accretion disk exhibits a quick adjustment in its behavior and the accretion rate. This variability is generated by the accretion flow, and, in part, by the turbulence in this accretion flow, and can have a timescale $\gtorder R_{\rm SMS}/v_{\rm r} \sim 3\times 10^{13}\,{\rm cm}/(10-100\,{\rm km\,s^{-1}})\sim 0.1-1$\,yr, as confirmed by Figure\,\ref{fig:time_rate}. Of course, the dynamics of accretion flow within the parent DM halo, larger turbulent eddies, and the cosmological conditions can impose longer timescales compared to this short timescale.

We have estimated the duty cycle, which is defined as the fraction of the time when accretion is not being suppressed, as $\sim 0.9$. This means that the radiation feedback does not completely stop the sink growth, probably due to the disk accretion being dominant in the vicinity of the sink, and having a highly anisotropic 3-D character.

Previous works, which deal with accretion onto the SMBH, have argued that the accretion rate is strongly suppressed and cannot keep the high rate. They have questioned the possibility of the seed BH growing to SMBH mass at $z=7$ \citep[e.g.,][]{johnson11,latif18,regan19}. In our simulation, we observe that the accretion rate onto the sink drops well below the Eddington rate but for a very limited time period. The accretion resumes again in an episodic way.  The explanation for this difference lies in that our radiation feedback is weaker than in the case of the SMBH, and because our accretion is strongly anisotropic, the radiation is capable of escaping in the preferred directions. This result is more realistic as it follows from the 3-D radiation transfer.

\begin{figure*}
\center
\includegraphics[width=0.99\textwidth,height=0.15\textheight,angle=0]{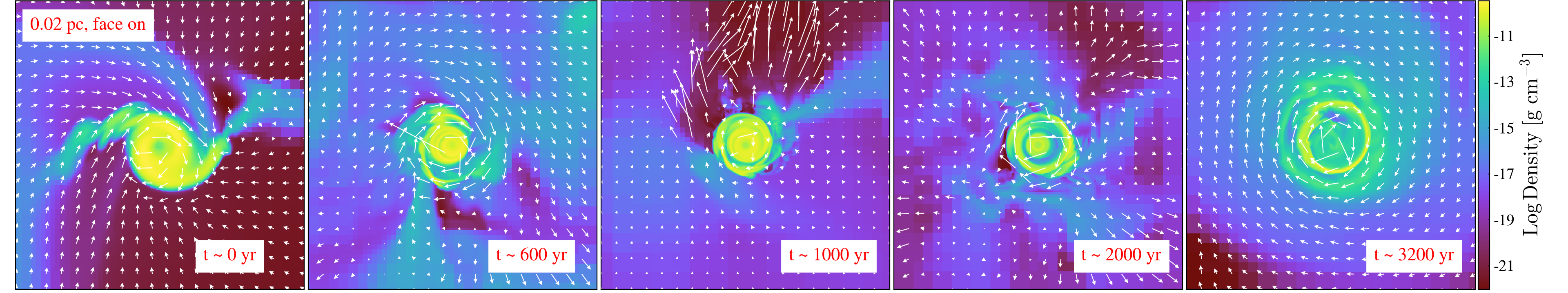}
\includegraphics[width=0.99\textwidth,height=0.15\textheight,angle=0]{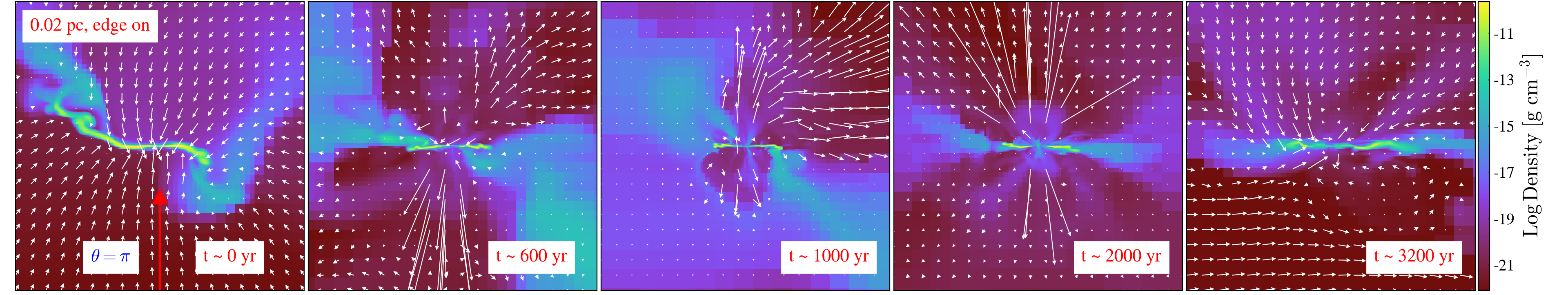}
\includegraphics[width=0.99\textwidth,height=0.15\textheight,angle=0]{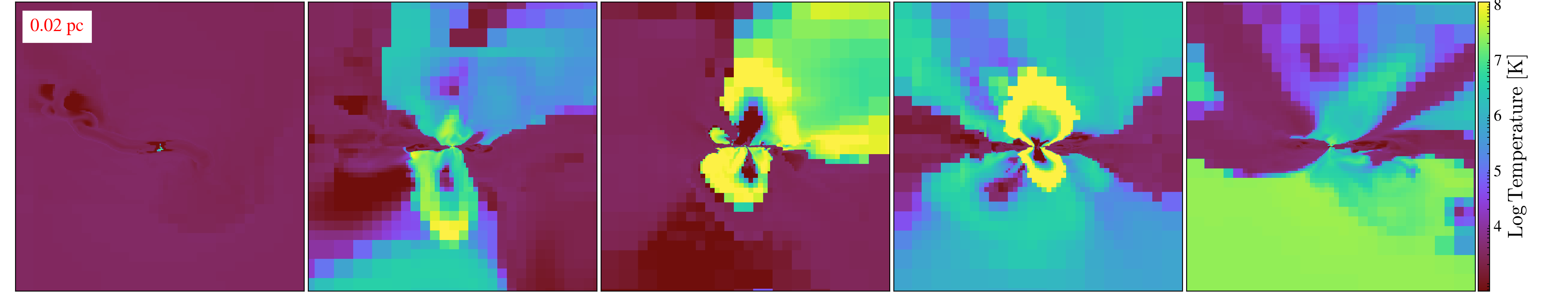}
\includegraphics[width=0.99\textwidth,height=0.15\textheight,angle=0]{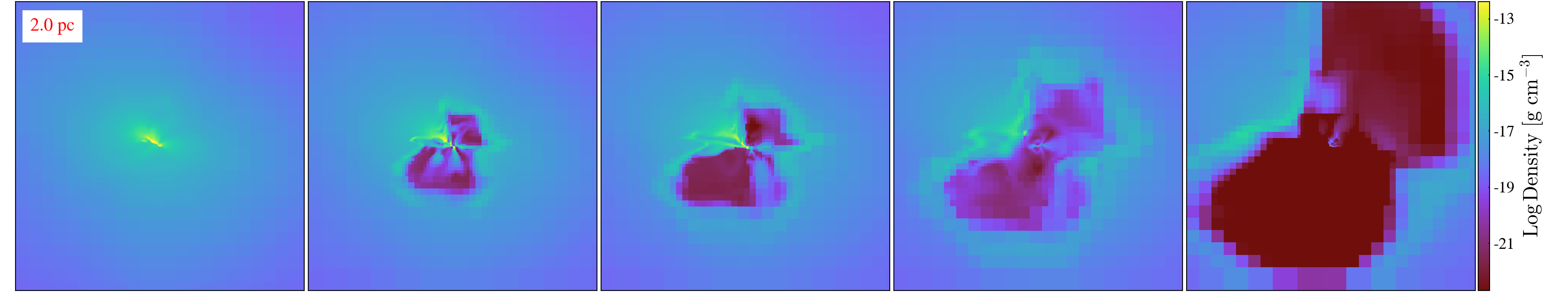}
\includegraphics[width=0.99\textwidth,height=0.15\textheight,angle=0]{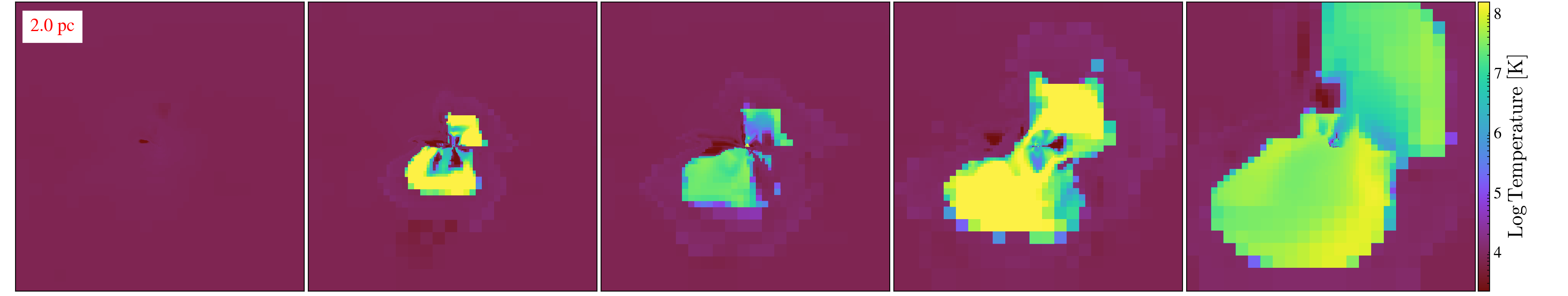}
\caption{Evolution of the outflow from the sink particle and the surrounding accretion disk driven by the radiation feedback. The slice snapshots are shown on the scale of $2 \times 10^{-2}$\,pc and $2.0$\,pc. {\it Top row:} the arrows indicate the direction and the value of the outflow velocities. {\it Second row:} From left to right, an anisotropic outflow is generated above and below the disk. The long red arrow in the panel at $t=0$ points to the position of the sink. At $t=0$, the outflow is not yet visible and the accretion dominates. The $t=1,000$\,yr snapshot displays a developing outflow in the upper hemisphere and a growing cavity there. At $t=2,000$\,yr, the outflow is substantially collimated and pushes the gas into an expanding shell, outlining the low-density cavity. The $t=3,000$\,yr snapshot shows the biconical outflow, which has expanded beyond the disk. The upper and lower bubbles are asymmetric. Note the dominance of the outflow and generated bubble below the accretion disk.}
\label{fig:outflow}
\end{figure*}

\begin{figure}
\center
\includegraphics[width=0.45\textwidth,height=0.27\textheight,angle=0]{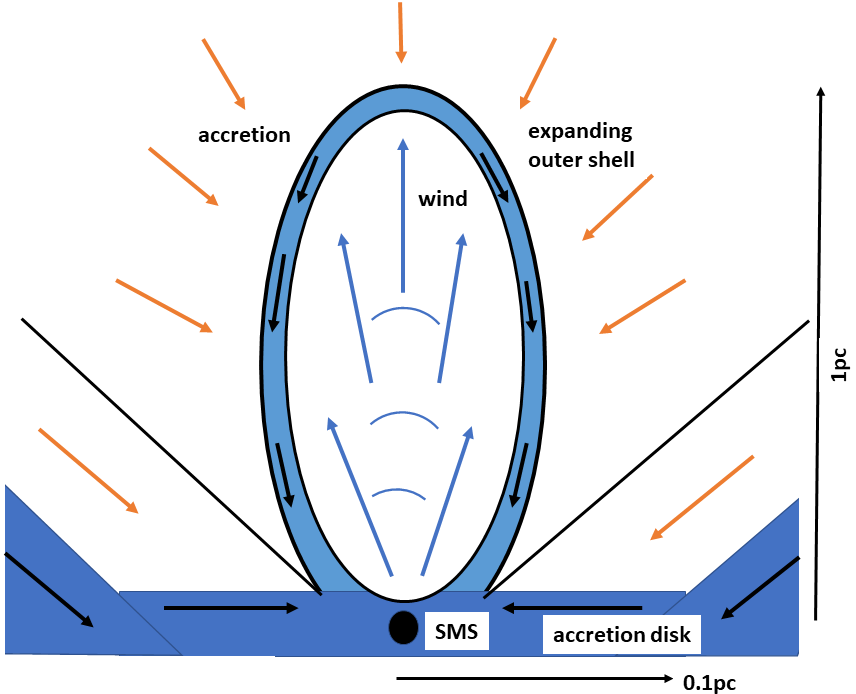}
\caption{Sketch of the inflow/outflow simulated in Figure\,\ref{fig:outflow}:  expanding outer shell forming a hot bubble filled with the fast wind and shells formed by injections of the accretion material, enveloped by the ambient accretion flow and accretion disk. The shocked ambient accretion and the fast wind backflow onto the outer shell at $t\sim 3\times 10^3$\,yr. The expanding cavity is transformed into the funnel. In Fig.\,\ref{fig:outflow}, the bubble is expanding to the lower side of the accretion disk. Only one hemisphere is shown here.}
\label{fig:sketch}
\end{figure}

\subsection{Accretion disk formation}
\label{sec:b_disk}

With the growth of the sink, its gravity dominates and the gravitational torques, which play a large role in extracting the angular momentum from the gas become inefficient in the vicinity of the sink. In fact, this is the main reason for weakening the extraction of angular momentum from the accretion flow and disk formation around the sink. We observe and follow the disk growth after the sink mass exceeds $\sim 10^3\,M_\odot$.  As we describe below, the disk mass is about a few percent of the sink mass, and this fraction never increases above $\sim 5\%$ during our simulation, either before or after triggering the radiation feedback.

Figure\,\ref{fig:disk} displays the properties of this disk at $t\ge 0$ in cylindrical shells at different times. We determine the extent of the disk by calculating the ratio of the specific angular momentum of the gas, $j$, to the Keplerian specific angular momentum, $j_{\rm K}$, and define the disk by $j/j_{\rm K}\gtorder 0.4$.

The disk thickness is constant, $h\sim 7\times 10^{-4}$\,pc, within its approximate radius of $\sim 10^{-2}$\,pc, and this thickness increases linearly beyond this radius with an opening angle of $\sim 20^\circ-30^\circ$ till $\sim 1$\,pc, where it merges with the accretion flow. Our simulation comfortably resolves the vertical structure of the disk. Initially, the disk surface density, $\Sigma(r)$ displays a central flat core within $r\sim 2\times 10^{-3}$\,pc, followed by $\Sigma(r)\sim r^{-2}$ dependency inside $\sim 1$\,pc. Between $t\sim 2,000$\,yr and $\sim 3,200$\,yr, a hot cavity has developed between $r\sim 2\times 10^{-3}$\,pc and a  few$\times 0.1$\,pc, followed by a dense shell around 1\,pc. At this time, the disk mass fraction of the sink falls below 1\%. The typical size of accretion disk is $\sim 0.01$\,pc, where the disk is supported predominantly by rotation.

The disk never exhibits any sign of fragmentation due to its local gravity, $\sim \pi G\Sigma(r)$. The reason for this is clear --- the disk is stabilized by the vertical component of the sink particle gravitational acceleration, $\sim GM_{\rm sink}(h/r)/(r^2+h^2)$. At all radii within $\sim 1$\,pc, the sink particle dominates. Note, that additional components, such as the DM halo contribution and turbulence in the disk \citep{begelman09,choi13} will contribute to the local stability of the disk against fragmentation. Nevertheless, we have calculated the Toomre's $Q(r)$ parameter for various times of evolution \citep{toomre64}, and found that it is always greater than a few, even when only the sound speed of the gas in the disk is used to calculate $Q$. When we invoke the turbulent velocities in the disk, the minimum of $Q$ lies around 10. This confirms that our disk will not fragment.

The size of the disk fluctuates wildly outside $\sim 10^{-2}$\,pc. This variability is clearly related to the feeding plane of the accretion flow coming from larger radii, as well as to the expanding hot cavity generated by the feedback. The feeding plane can be inclined by an angle up to $\sim 30^\circ$ compared to the plane of the inner disk and correlates with the prevailing angular momentum in the inner accretion flow.

Because the shear in such a disk is low, the viscous heating is negligible and its temperature remains close to the floor. The high column density leads to a low temperature of $\sim 10^4$\,K, except after $t\sim 2,500$\,yr, when the {\it extended} disk is basically destroyed and the temperature rises above $\sim 10^6$\,K. Hence, the inner disk luminosity can be neglected in the simulation of the feedback as its effective temperature will peak below the UV range.

Shown in Figure\,\ref{fig:disk}, the inflow velocity inside the disk is $\sim 10-100\,{\rm km\,s^{-1}}$, oscillating in the vicinity of the sink. Around $t\sim 2,000$\,yr, the outer disk outside $\sim 10^{-2}$\,pc is destroyed, but is gradually restored towards $t\sim 3,200$\,yr. The accretion rate has been calculated using cylindrical radii, and reflects the destruction of the outer disk. This destruction results from the expansion of the hot cavity generated by outflow. It can be seen directly in the temperature panel of Figure\,\ref{fig:disk}.

When plotted along specific rays, the density, velocity and temperature profiles at $t=0$ display the anisotropy of density and temperature distributions around the sink. Considering the accretion disk midplane and its rotation axis, we use the polar angle $\theta$ to specify different rays. The angle $\theta=0$ points along the upper part of the rotation axis, while $\theta=\pi$ along the lower part of the axis, thus dividing the volume into the upper and lower hemispheres. As evident from the properties of the inflow-outflow cycle discussed in the subsequent sections, they differ in both hemispheres, and even display aspect angle dependence within each hemisphere. For example, the density of the accreting material is higher in the upper hemisphere by a factor of a few, and this results in asymmetric outflow properties there. 

To follow the density and temperature behavior inside the expanding cavities in both hemispheres, we focus on two rays, $\theta = \pi/4$ and $\theta=\pi$. The reason for this choice becomes obvious at later times in Figure\,\ref{fig:outflow}.

\begin{figure*}
\center
\includegraphics[width=0.8\textwidth,height=0.7\textheight,angle=0]{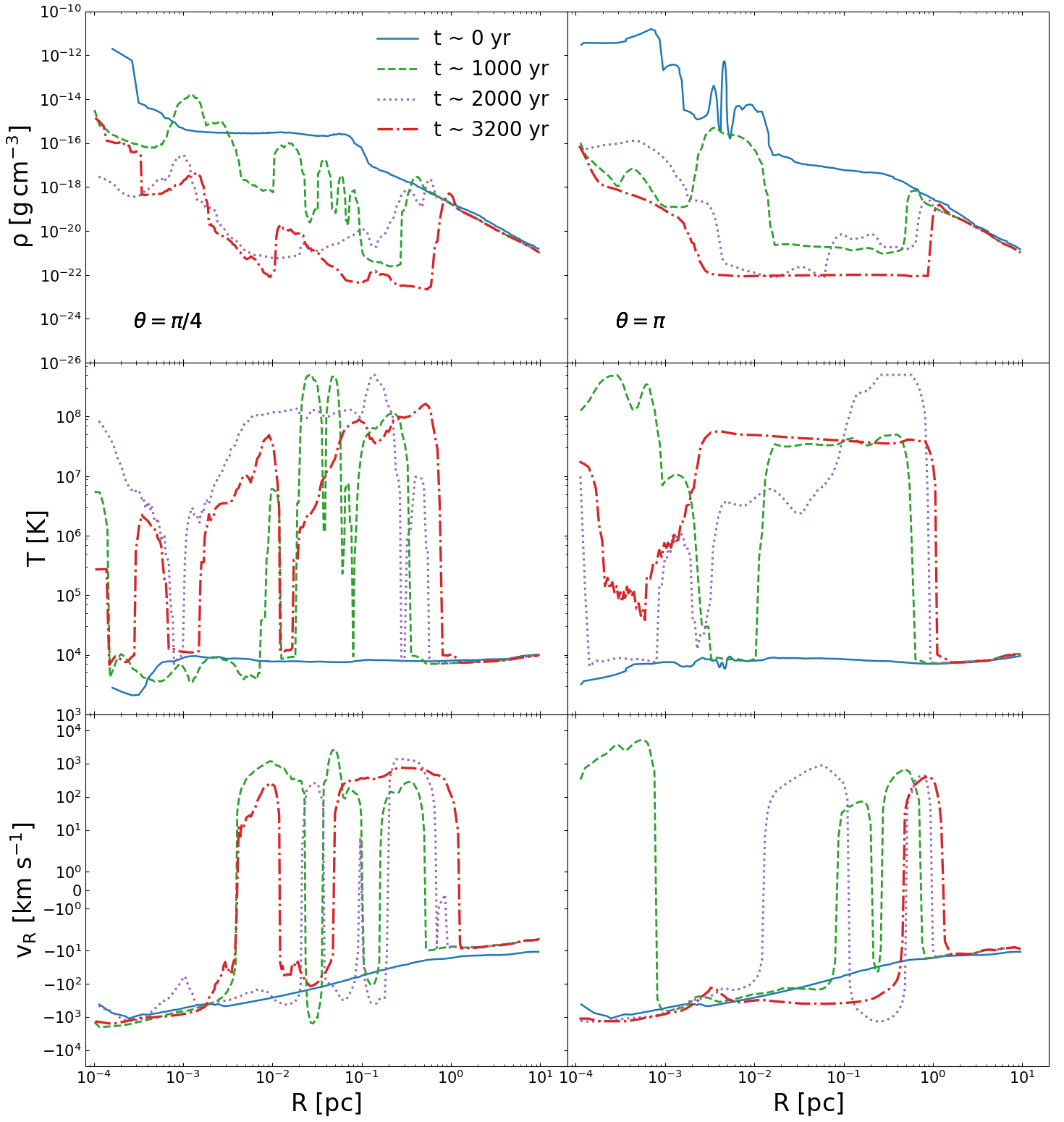}

\caption{Radial profiles of density $\rho$ (top frame), temperature $T$ (middle frame), and inflow/outflow velocity $v_{\rm R}$ (bottom frame) along the ray $\theta=\pi/4$ (left column) and $\theta=\pi$ (right column) coinciding with the disk rotation axis in the lower hemisphere. The times correspond to the snapshots shown in Figure\,\ref{fig:outflow}. Note that positive velocities represent the outflow and the negative ones represent the inflow.}
\label{fig:profile}
\end{figure*}

\subsection{The radiation feedback and the outflow evolution}
\label{sec:outflow}

The typical velocity of the accretion flow onto the sink varies in the range $10-10^3\,{\rm km\,s^{-1}}$, depending on whether it comes in a quasi-spherical or disky fashion. Kinetic accretion energy is thermalized and produces the blackbody emission with the effective temperature of $\sim {\rm few}\, 10^4$\,K, emitting mostly the ionizing UV photons and some soft X-rays. Hence, the inescapable corollary of accretion onto the sink particle representing the central entity, either SMS or a self-gravitating disk, in the presence of the radiation feedback, is the development of a wind from the vicinity of the sink. The internally-generated thermonuclear contribution to the luminosity is less than 50\% of the accretion luminosity, unless the latter dips due to the decreased mass accretion rate.

To understand the effect of the radiation/thermal feedback, we turn to the radiation transfer. The sink is embedded within the accretion disk, with the disk being substantially geometrically thicker than the sink size. So the radiation is initially trapped in the center. The gas density surrounding the sink at $t=0$ is around $\sim 10^{-12}\,{\rm g\,cm^{-3}}$, but this average is misleading. Some of the gas is at a much lower density and, therefore, could be completely ionized. Higher-density gas which is injected by the inflowing material, has a shorter recombination timescale --- this gas remains neutral. With the radiation and thermal energy accumulated in the central region, the radiation and thermal pressure gradients exert outward forces and start to clean the gas around the sink, preferentially in the direction along the disk rotation axis.

Figure\,\ref{fig:outflow} displays the density and temperature slices of this outflow with a superposed velocity field at different times and on different scales. For clarity, we display the snapshots with the face-on and edge-on accretion disk. 

However, a steady state is never achieved in the center, as cold gas injections are sporadic. After a short transient phase, hot bubbles, i.e., cavities, start to expand above and below the disk plane. The bubbles are delineated by dense expanding shells which accumulate the material from the bubble interior as well as from the exterior injected accreting gas. Because the density profiles of the accreting material differ above and below the disk plane, as we have discussed earlier, the outflow properties differ as well. The bubble above the midplane develops first, but the density below the disk plane is lower, hence the expanding bubble there grows faster.  

By $t\sim 1,000$\,yr, the inner spinning disk and the upper and lower hemisphere outflows are fully developed (see the upper two rows of Figure\,\ref{fig:outflow}). It is anisotropic and collimated within the angle of $\sim 30^\circ-50^\circ$. The flow asymmetry has increased: below the disk, the cavity has expanded to $\sim 0.5$\,pc, while above the disk, the cavity size is smaller, as seen in the third row of this Figure. 

By this time both bubbles become deformed. The radiation acceleration is lower than the thermal acceleration, $a_{\rm rad}/a_{\rm P}\sim 10^{-1}$ inside the bubbles, as we show later. But this ratio increases dramatically with the approach of the expanding shell and becomes $\sim 1$ inside the shell. We also observe that the inner part of the disk has been truncated, in agreement with Figure\,\ref{fig:disk}.  

At $t\sim 2,000$\,yr, the upper bubble is about a factor of $\sim 2$ smaller than the lower one, and the asymmetry and skewness of the outflow are obvious.  Because the gas density in the cavity close to the disk is higher, this gas cools faster. On a scale of 2\,pc (fourth row), the disky inflow plane becomes inclined to the innermost disk --- this happens to the change in the angular momentum vector of the accretion flow. 

Note the high temperature in both bubbles, which has reached $\sim 10^8$\,K, and the very low temperature inside the disk, $\sim 8,000$\,K. The ratio $a_{\rm rad}/a_{\rm P}$ follows the expanding shells, where it is of the order of unity.

By $t\sim 3,200$\,yr, the bubbles in the fourth  row of Figure\,\ref{fig:outflow} continue to expand with $v_{\rm s}\sim 200\,{\rm km\,s^{-1}}$. We distinguish the velocities of the dense shells, $v_{\rm sh}$, from that of the fast wind within the cavities, $v_{\rm w}$. The wind velocity inside the cavities, whose densities become progressively lower and reach $\sim 10^{-21}\,{\rm g\,cm^{-3}}$, varies up to $v_{\rm w}\sim {\rm few}\times 10^3\,{\rm km\,s^{-1}}$. The temperature inside the cavities varies between $\sim 10^4-10^8$\,K (Fig.\,\ref{fig:profile}). The lower limit on $T$ results from the injection of the dense and cold material from the disk, as discussed below. As before, the radiation force is comparable to the thermal pressure gradients inside the shells.

\subsection{The bubble expansion}
\label{sec:bubbles}

We observe multiple dense shells which form as a result of the cold material injection by the accretion flow to the region above and below the disk. Figure\,\ref{fig:sketch} provides the sketch of the inflow and outflows interaction driven by the radiation force, formation of the dense outer shell, and the backflow of the shocked ambient accretion and shocked wind. These shells move faster than the outermost shell, and are recognizable by high density and low temperature in Figure\,\ref{fig:profile}. They approach and merge with the outer shell, which has $T\sim 10^4$\,K. In the outer shell, we do not resolve between the post-shock ambient accreting matter and that of the shocked wind, as these mix quickly. Figure\,\ref{fig:profile} displays the properties of the outflow along the ray with $\theta=\pi$. Initially, the bubble is strongly elongated, but by $t = 3.2\times 10^3$\,yr, it became more round, which points to the dominating role of thermal pressure gradients compared the radiation force. Note the very shallow, nearly constant density profile in the cavity, forming a broad plateau.

The accreting material appears to be injected into the cavities sporadically and can be distinguished by the low temperature and high density initially, and additional gas is ablated by the radiation hitting the accretion disk and its surrounding disky accretion flow. We observe the mass loading of the wind inside the cavities. The injected material, when accelerated, is squeezed into thin shells as it expands. The velocities of these shells are much faster than that of the outermost shell. Hence they quickly reach the outer shell and contribute to the column density there. 

As the accretion flow produces an azimuthally-distributed and inclined wedge around the sink and its accretion disk, some of the accretion inflow will be ablated radiatively and hydrodynamically. The surface area of the accretion flow increases with radius and will contribute progressively more to the mass-loaded wind --- indeed, the outflow rate increases with the distance from the sink, as we show below.  

The injected and ablated material from the accretion flow both contribute to the outflow. Because the outflow is very anisotropic, and because of the mass loading by the outflow, we have estimated its rate by selecting cells with positive radial velocities within the limited range of the polar angles in the upper and lower hemispheres for $t = 3.2\times 10^3$\,yr. The limit on the polar angle has been used in order to separate the inflow from the outflow. To avoid contamination with the disk accretion, we only show the accretion rate for radii larger than the accretion disk. Accretion rate within the disk only has been shown in Figure\,\ref{fig:disk}.

The structure of the cavities is complex. Some of the material that is injected into the cavities is not accelerated sufficiently and falls back onto the disk at larger radii, i.e., it is recycled back. The other complication comes from the fact that the bubble is not exactly spherical and is not expanding spherically, but is elongated, as it expands preferentially along the path of the least resistance, which varies with time. Lastly, the bubble appears to be leaky --- its column density along the rays close to the disk rotation axis quickly achieves $\sim 10^{23}\,{\rm cm^{-2}}$ and does not change over the run time. As we show in Section\,\ref{sec:extrapolate}, the material flows backward along the outer shell. So the column density does not increase as expected in the snowplow phase. The mass loading of the wind is difficult to quantify, as the gas is injected sporadically, and some of the gas is pushed sideways.

These properties of the bubble and its confining shell differ from the typical evolution of the expanding spherical shells \citep[e.g.,][]{faucher-giguere12}. In general, the physical conditions of the outflow differ substantially from that in the AGN because of the near absence of the Compton heating and inverse Compton cooling, as the radiation spectrum in the SMS accretion disk case is substantially milder --- a blackbody compared to the hard power law of AGN. Lastly, the wind velocities in the SMS-driven outflow, $\ltorder 10^{2-3}\,{\rm km\,s^{-1}}$, are smaller than those expected in the AGN, and the post-shocked material is expected to have the same temperature for protons and electrons, unlike much faster AGN winds. Nevertheless, these fast shocks propagating within the cavity provide a powerful heating mechanism of the gas, even to $t\sim 10^8$\,K.

The cooling timescale of the wind inside the cavities, $t_{\rm w,cool}$, when they expand beyond $\sim 10^{-2}$\,pc, is very long, $t_{\rm w,cool}\sim 3\times 10^4(R/0.01\,{\rm pc})^2$\,yr. The wind crossing time of the cavities is $t_{\rm w,cross}\sim 3 (R/0.01\,{\rm pc})$\,yr. Hence $t_{\rm w,cool} >> t_{\rm w,cross}$, and the wind behaves adiabatically within the cavities. This estimate was performed for $\theta = \pi/4$ and $\pi$. The wind velocity is somewhat slower for larger inclinations, but this inequality still holds.

Next, we compare the cooling time of the shocks (i.e., fast-moving shells) inside the cavities with the shock crossing time to reach the outermost shell. The cooling time is $t_{\rm cool}\sim 30$\,yr, while  the crossing time is $t_{\rm cross}\sim {\rm few}\times 10^3$\,yr inside 1\,pc. This makes both of them much shorter than the expansion timescale of the outer shell,  i.e., $t_{\rm s}\sim 2\times 10^4$\,yr. Hence, we cannot assume that the shocks inside the cavities remain adiabatic.

\begin{figure*}
\center
\includegraphics[width=0.98\textwidth,height=0.15\textheight,angle=0]{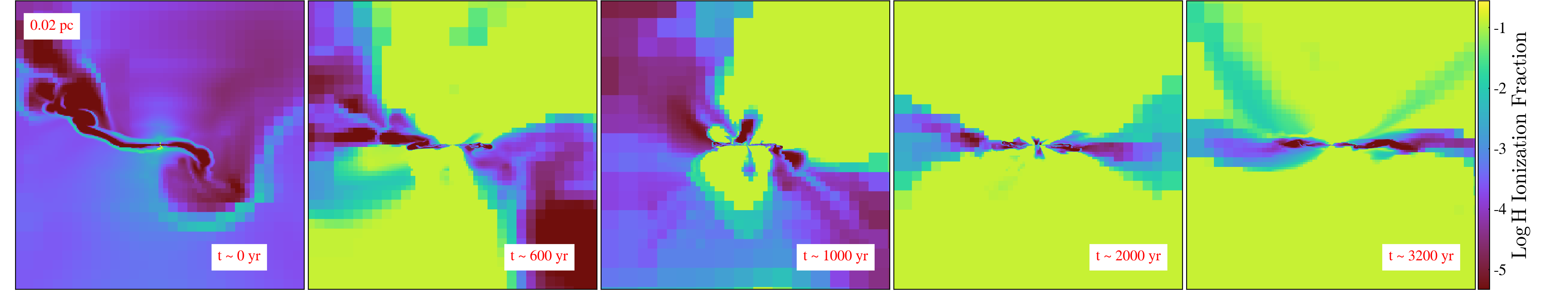}
\includegraphics[width=0.98\textwidth,height=0.15\textheight,angle=0]{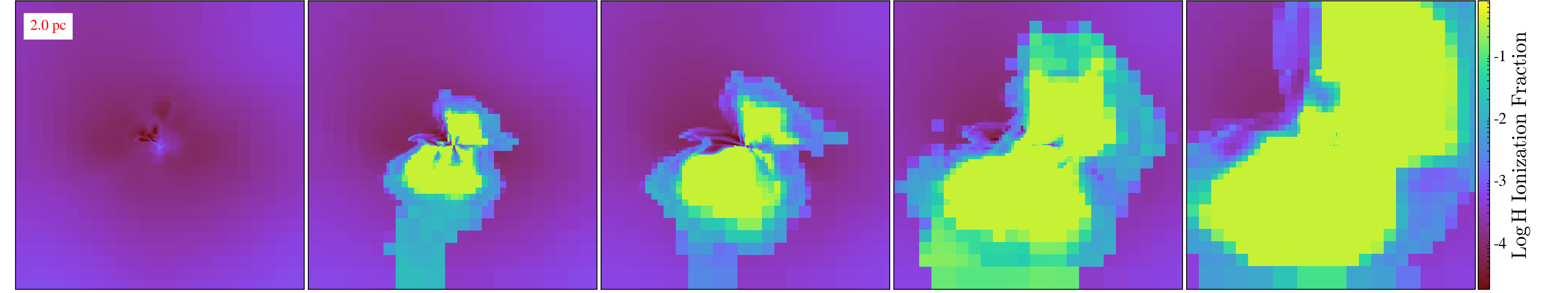}
\includegraphics[width=0.98\textwidth,height=0.15\textheight,angle=0]{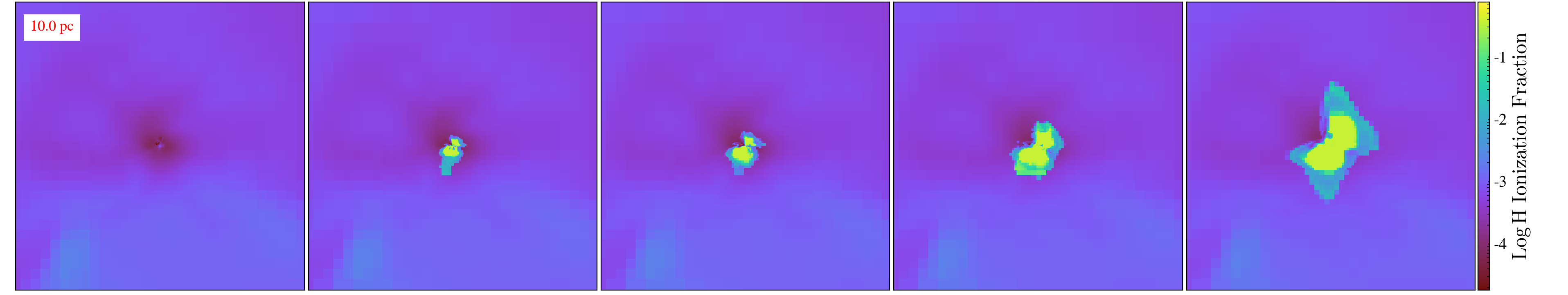}
\includegraphics[width=0.98\textwidth,height=0.15\textheight,angle=0]{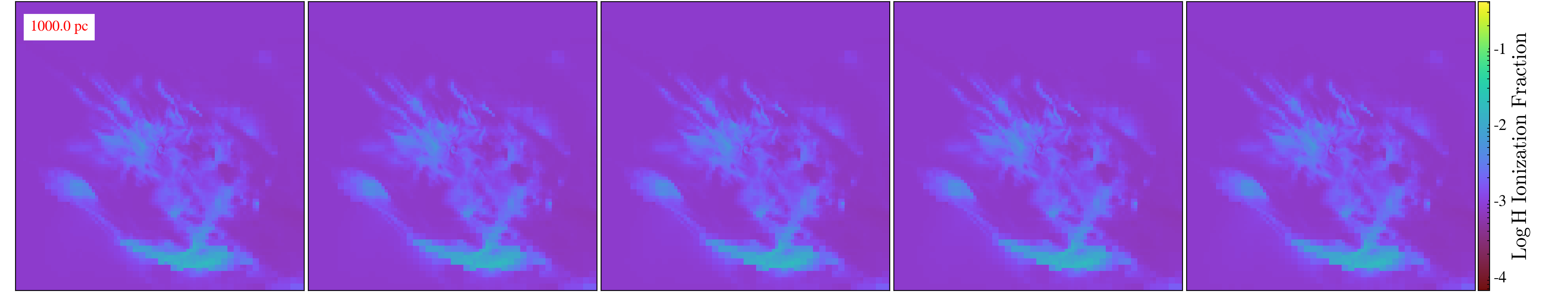}
\caption{Evolution of the ionization cones in the inflow/outflow within the DM halo hosting a sink particle represented by the SMS or a self-gravitating accretion disk. The outflow is driven by the radiation and thermal feedback. The slice snapshots are shown on the scale of 0.02\,pc (top), to 1000\,pc (bottom), with the accretion disk situated edge-on in all frames. At $t=0$, the outflow does not exist, and the accretion dominates. Note the accretion shock at $\sim 500$\,pc which persists for the entire simulation (bottom frames). At $t=0$, the outflow does not exist, and the accretion dominates. From $t\gtorder 600$\,yr, the snapshots show a biconical ionization structure which is visible up to $\sim 2$\,pc. More in the main text. }
\label{fig:ioniz}
\end{figure*}

\subsection{Ionization state of the outflow}
\label{sec:ionization}

As stated earlier, the outflow from the sink region is anisotropic, and it develops two cavities which grow into the funnel shape. This funnel forms in response to the radiation and thermal pressure gradients and is partially collimated by the inflow material. The ionization field is biconical due to the obscuration close to the sink and because of the geometry of the accretion flow. 

During the simulation time of $3.2\times 10^3$\,yr, the photons can in principle reach $\sim 1$\,kpc, and the radiation can affect the entire DM halo. However, most of the radiation flux is absorbed by the expanding shells. Therefore, we have analyzed the ionization balance within the host DM halo. Before the radiation feedback has been initiated, the gas stays nearly neutral, and the highest level of ionization, $\sim 10^{-2}$, corresponds to the standing shock just inside the virial radius, as shown at $t=0$ for 1\,kpc frame in Figure\,\ref{fig:ioniz}. Elsewhere, in the accretion flow, the ionization fraction is $\sim 10^{-3}$, and the gas temperature stays at the floor of the neutral hydrogen. 

\begin{figure}
\includegraphics[width=0.45\textwidth,height=0.3\textheight,angle=0]{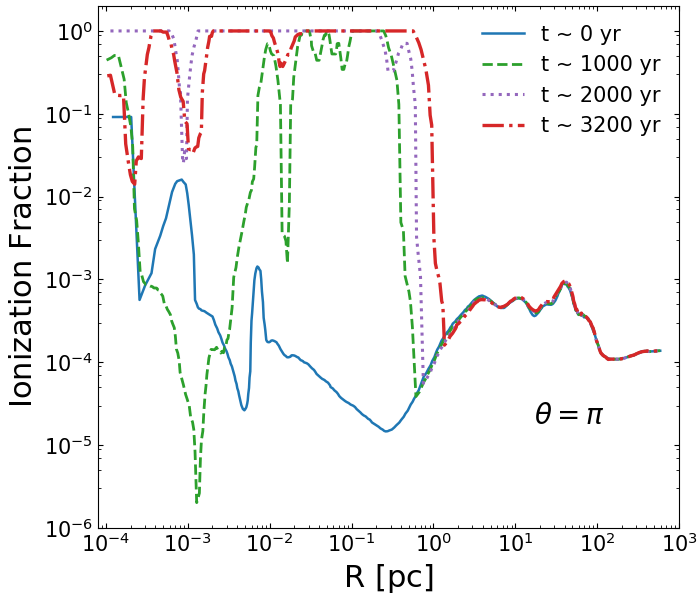}
\caption{Ionization fraction profiles measured along the $\theta=\pi$ ray at few representative times.}
\label{fig:ioniz_profile}
\end{figure}

By $t\sim 600$\,yr, the outer shell has reached $R_{\rm s}\sim 0.3-0.5$\,pc, depending on the direction. Even inside the bubbles, the gas is not completely ionized because of cold gas injection and the formation of fast-moving shells (e.g., Figure\,\ref{fig:ioniz_profile}). In the very vicinity of the sink, the ionization level is lower due to the high density and low temperature of the injected accretion gas into the funnel. This situation changes with time only due to the outer shell propagating to larger radii.

At $t = 3.2\times 10^3$\,yr, the ionization region has reached $R_{\rm ion}\sim 1.3$\,pc along $\theta = 0$ and $\theta\sim \pi$ rays. The outer shell remains weakly ionized at all times, at $\sim 10^{-2}$ (e.g., Figure\,\ref{fig:ioniz}). The ionization has a biconical geometry in both hemispheres. The opening angle of the ionization cones, determined from 0.1\,pc snapshots of gas distribution and which is edged by the outer shell in both hemispheres, is $\theta\sim \pi/4$. 

The difference between the gas densities along the polar axes and the equatorial plane has increased with time and, by the end of the simulation, this ratio is $\rho(\theta=\pi)/\rho(\theta=\pi/2)\sim 10^{-3}-10^{-4}$. The polar axes have been practically emptied by this time, while the accretion disk midplane is very dense. However, the disk itself is clearly sandwiched by the accretion wedges containing inflowing gas with a high angular momentum --- this gas is the main contributor to the obscuration and forms the edges of the ionization cone. The gas is highly ionized within the cones, inside the radii of the outer shell, $\sim 1.3$\,pc. Between this radius and $\sim 10$\,pc, the ionization fraction is $\sim 10^{-3}-10^{-4}$. 

We also note that the mass supply to the innermost region has been disrupted during $\sim 2,000-3,000$\,yr and becomes intermittent during this time period as can be observed in Figures\,\ref{fig:time_rate}. Figure\,\ref{fig:disk} also shows that the outer disk which extends through this region is largely destroyed for about$\sim 10^3$\,yrs.

\subsection{Sub-parsec evolution of rotation}
\label{sec:rotation}

\begin{figure}
\includegraphics[width=0.45\textwidth,height=0.3\textheight,angle=0]{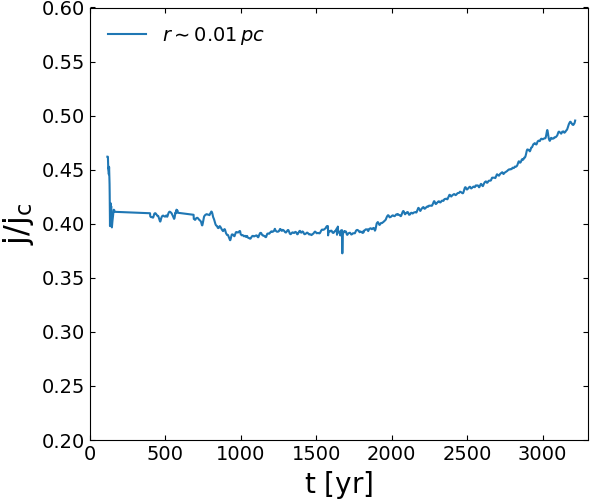}
\caption{Evolution of the specific angular momentum of accreting gas on a scale of 0.01\,pc normalized by the circular specific angular momentum.}
\label{fig:j-ratio}
\end{figure}

The accretion flow modeled here has its seed angular momentum similar to those of parent DM halos, and is negligible at larger radii. However, the gas in direct collapse accumulating in the central parsec has its angular momentum substantially affecting its dynamics. This is also reflected in the formation of an accretion disk on these scales.  Figure\,\ref{fig:j-ratio} displays the evolution of $j/j_{\rm c}$ --- the ratio of the specific angular momentum to that of circular specific angular momentum within 0.01\,pc. The circular angular momentum has been calculated using the tangential velocity of $v_\phi^2 = GM(<r)/r$, where $M(<r)$ is the total mass within $r$. At $t=0$, the gas motion within these regions is not affected by the outflow and by radiation and thermal feedback, and is high, above 0.5. As the outflow is generated, the angular momentum shows some gradual decrease, especially on smaller scales. Then increases again and represents a rotationally supported accretion disk. 

The high fraction of angular momentum on sub-parsec scales has a direct effect on the forming central mass concentration. Although in this work, we do not resolve the scales inside $10^{-5}$\,pc, one would expect that the growing mass there will have a comparable angular momentum. This has been confirmed by simulations without the sink particles which resolved these scales \citep{luo18,ardaneh18}, but have been left unpublished.

\section{Discussion}
\label{sec:discuss}

We have studied the effect of radiative and thermal feedback from growing central mass accumulation during direct collapse within the parent DM halos at high redshift, in the form of an SMS or a self-gravitating disk. We performed 3-D cosmological simulations and modeled the emerging radiation flux with a ray-tracing algorithm on the fly. The accretion rate on a large scale is determined by the DM halo potential, while it is modified on the sub-pc scale when the central mass approaches $\sim 10^5\,M_\odot$, modeled as a sink particle with a radius of $\sim 1.3\times 10^{-5}$\,pc. 

We analyze our results by summarizing them first.
\begin{itemize}
    \item The accretion rate onto the sink is very noisy, in part because of the natural variation in the accretion rate, and in part because of the radiation/thermal feedback.  Accretion onto the sink is mildly super-critical or critical, with the duty cycle of $\sim 0.9$, when the rate drops substantially. The emerging luminosity mimics the evolution of the accretion rate and peaks at $\sim 2L_{\rm Edd}$. The anisotropy of the accretion flow generated by the presence of the angular momentum appears to sufficiently attenuate the UV and soft X-rays, which can help for the survival of the H$_2$ within the parent DM halo.
    
    \item While the accretion rate is independent of radius on the DM halo scales, it is slowly increasing inwards inside 0.01\,pc scales acquiring a disky character when the massive object forms. The outflow is generated by the feedback onto the accretion flow, and is characterized by redirecting some of the accretion flow at the smallest radii and ablating the accretion from the boundaries of the biconical funnel. The dominant opacity of the primordial composition gas is found to be the electron scattering and the bound-free hydrogen absorption. 
    
    \item Accretion disk of $\sim 0.1$\,pc forms around the sink when it reaches a mass of $\sim 10^3\,M_\odot$. No fragmentation has been observed in the disk. The outflow has been triggered by the radiation/thermal feedback, which deposits momentum and energy within the biconical region around the rotation axis. Although the outflow is biconical, its geometry and evolution differ substantially in the upper and lower hemispheres with respect to the disk plane. We emphasize that no constraints have been imposed on the symmetry of the accretion and outflow ---  the evolution of the inflow/outflow depends sensitively on this point.

    \item Hot cavities form and continue to grow, with the interior gas being sporadically injected into the biconical funnel, accelerated and compressed into thin shells which propagate outwards and merge with the outermost shells. Overall, these cavities reflect the formation of the biconical funnel along the rotation axis which is substantially depopulated of gas.
    
    \item The structure of the growing funnel is complex and asymmetric. We follow the evolution of the best-developed funnels in both hemispheres. The density profile within each funnel is very shallow with $\sim 10^2-10^3\,{\rm cm^{-3}}$, and the temperature is $\sim 10^{7-8}$\,K.
    
    \item The wind velocities in the funnels vary up to ${\rm few}\times 10^3\,{\rm km\,s^{-1}}$, while the outer shells have decelerated to $\sim 200\,{\rm km\,s^{-1}}$ by the end of the run and has crossed $\sim 1.3$\,pc at $t=3.2\times 10^3$\,yr.
    
\end{itemize}

Next, we aim to understand whether the outer shell edging the funnel will be further decelerated, or it will be able to break out of the central parsec and expand to the scale of  few\,$\times 100$\,pc. The final result has strong implications for escaping radiation, continuum and Ly$\alpha$ from these objects, and their detectability.

\begin{figure}
\center
\includegraphics[width=0.45\textwidth,height=0.6\textheight,angle=0]{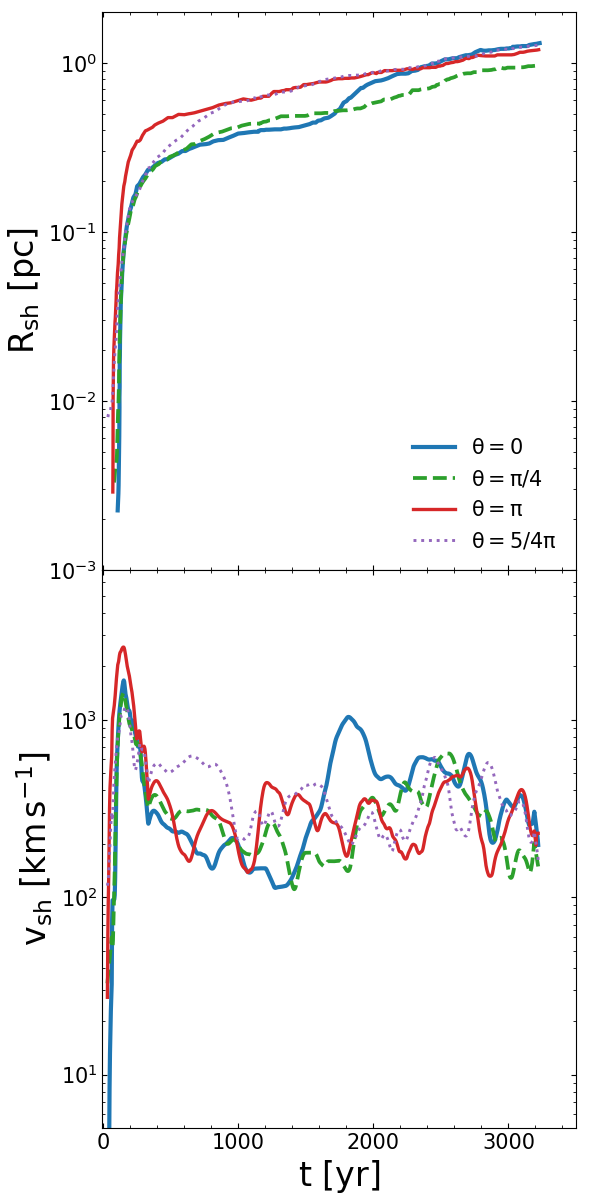}
\caption{Top: radius of the expanding shells, $R_{\rm sh}$, as a function of time in different inclinations above and below the accretion disk: $\theta=0$ ray points along the rotation axis of the disk in the upper hemisphere, and $\theta=\pi$ points along the rotation axis in the opposite direction (Fig.\,\ref{fig:outflow}). Bottom: the expansion velocity of the shell, $v_{\rm sh}$, along the same directions. The outflow is asymmetric with respect to the disk midplane.}
\label{fig:shell}
\end{figure}

\begin{figure}
\includegraphics[width=0.45\textwidth,height=0.3\textheight,angle=0]{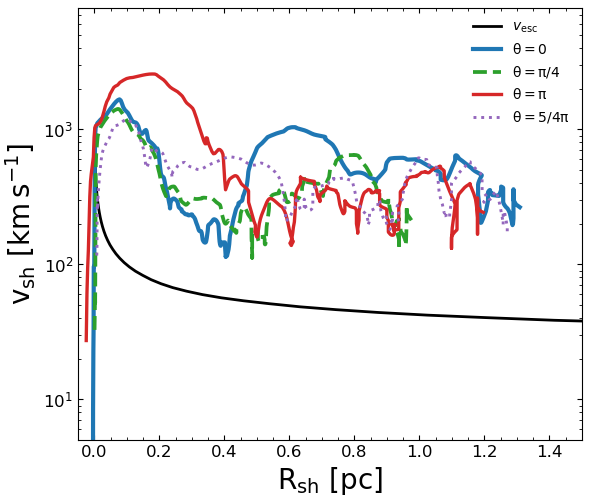}
\caption{Shell velocity, $v_{\rm sh}$, along various rays, as a function of the shell radius, $R_{\rm sh}$. The escape velocity from the region is shown for $M(<R_{\rm sh})$, which is the total mass within $R_{\rm sh}$, i.e., the gas, the sink, and the DM interior to the shell radius (black solid line).}
\label{fig:vs_vesc}
\end{figure}

\begin{figure}
\includegraphics[width=0.45\textwidth,height=0.6\textheight,angle=0]{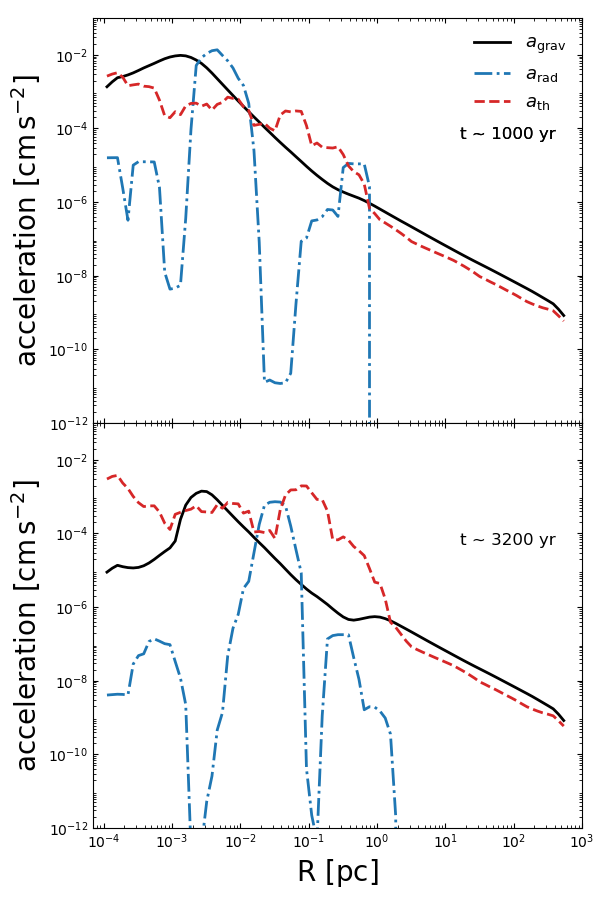}
\caption{Accelerations acting on the shell: thermal pressure gradient, radiation, and gravitational acceleration profiles as a function of radius, $R$, along the $\theta = \pi$ ray, at $t\sim 1,000$\,yr (upper frame) and $t = 3.2\times 10^3$\,yr (lower frame).}
\label{fig:acceleration}
\end{figure}

\subsection{The shell motion: breakout?}
\label{sec:extrapolate}

Figure\,\ref{fig:shell} provides us with the position and kinematics of the expanding shell in various directions. Our simulation has been continued through $t=3.2\times 10^3$\,yr after turning on the radiation feedback. At this time the shell is located at $R_{\rm sh}\sim 1.3$\,pc from the sink and expands with the velocity of $v_{\rm s}\sim 200\,{\rm km\,s^{-1}}$. The shell velocity has been decreasing from the peak of $\sim 2\times10^3\,{\rm km\,s^{-1}}$ at $t\sim 200$\,yr in the direction of $\theta=0$ to $\sim 100\,{\rm km\,s^{-1}}$ at $t\sim 1,300$\,yr, then rapidly accelerated again to $\sim 10^3\,{\rm km\,s^{-1}}$, and then decelerated by a factor of 3$-$5. These variations in the outflow velocity of the shell are both related to the radiation/thermal acceleration and variations in the accretion rate along the ray.

Note that after $t\sim 2,000$\,yr, the outer shell velocity along this ray is approximately flat in all directions, although the variability is non-negligible and there are clear differences in the expansion velocity along different rays. This confirms that the shell expansion is substantially anisotropic and depends on the complex interplay between the anisotropy of the driving force and the accretion flow. Such anisotropic expansion is also observed in supernova remnants \cite[e.g.,][]{williams18}. 

We have plotted the expansion velocity of the shell along four directions, including the $\theta = \pi$ ray, as a function of its radius, in Figure\,\ref{fig:vs_vesc}. These velocities have been superposed over the escape velocity from the same radii, based on the total mass enclosed within these radii. We stopped following the shell if its velocity dropped below the escape one. Note that the shell expands faster than the escape speed, basically always.

Most interestingly, while the shell expansion rate experienced periods of substantial slowdown and acceleration, the amplitude of these variations has gradually subsided as it approached 1\,pc, and it has basically leveled off. This statement can be applied to all directions in Figure\,\ref{fig:vs_vesc}.

What is the fate of this shell's expansion? Will the shell break through and escape from the DM halo, or will it decelerate and stop around $\sim 0.1-1$\,pc, as argued by \citet{regan19}? 

While continuing the simulation is prohibitive from the point of view of computational effort, we attempt to extrapolate the outflow kinematics and tackle its quenching or breakout options. Of course, the collimated AGN jet launched in the SMBH vicinity cannot be compared directly to the biconical wind from a similar mass but much less compact central object, e.g., the SMS or a self-gravitating accretion disk. But it is important to understand the wind effect on the accretion flow without the subgrid physics, as well as its broader implications.

At $R\sim 1$\,pc, the escape speed, $v_{\rm esc}\sim 38\,{\rm km\,s^{-1}}$. Hence the shell velocity is substantially higher than that of the escape velocity. The escape velocity depends on two factors, namely, the mass of the sink and the mass of the accreting gas. The latter contribution to the escape speed is a very weak function of $R$, because the gas density beyond the outer expanding shell, is about $\rho\sim R^{-2}$, and the gas mass dominates over the sink mass only outside $R\sim 5$\,pc, but at these distances it is the DM that becomes dominant.  So the escape speed tends to be constant.

At small radii, the shell is accelerated by the radiation pressure, but at larger distances from the sink, it is accelerated by the thermal pressure gradients, i.e., as along $\theta=\pi$ for example (Fig.\,\ref{fig:acceleration}). Why did the shell velocity become constant? Figure\,\ref{fig:acceleration} which displays the relevant acceleration profiles at $t\sim 10^3$\,yr and $t = 3.2\times 10^3$\,yr provides an insight.

First, we note that the total acceleration as a function of radius almost always exceeds the gravitational deceleration within 1\,pc. Second, the thermal acceleration almost always exceeds the radiation acceleration, except around the shell position at $t\sim 1,000$\,yr and around $\sim 10^{-2}-10^{-1}$\,pc at $t\sim 3,200$\,yr. This is especially visible inside the cavity at the latter time when the hot gas inside the bubble acts as a piston.  The temperature inside the bubble is $\sim 10^7$\,K at $t\sim 3.2\times 10^3$\,yr, but the temperature within the shell is $\sim {\rm few}\times 10^4$\,K, which drives the thermal gradient in pressure and accelerates the shell against the gravity.

The high radiative acceleration within $10^{-3}$\,pc is explained by a high opacity there --- this means high gas density in the vicinity of the sink and frequent injection of the cold accretion material there. The large acceleration around 1\,pc is explained by the opacity of the outer shell.  

Will the shell break out of the halo, or will it be confined to a $\sim 1$\,pc region? If the shell possesses velocity in excess of the escape speed, it can be decelerated by encountering the gas between the shell radius and $R_{\rm vir}$. Hence, as a next step, we compare the column density of the outer shell with the gas column density in the halo outside 1\,pc. We also compare the linear momentum, and kinetic energy per unit surface area of the shell at $t = 3.2\times 10^3$\,yr, with those of gas outside $R\sim 1$\,pc per unit surface.  If the shell maintains its velocity in excess of the escape speed from the region, it will break through the DM halo, unless it is decelerated by the gas column density. 

We estimate the shell column density, $N_{\rm sh}$, from its thickness $\Delta R_{\rm sh}\sim 0.1 R_{\rm sh}$, where $R_{\rm sh}$ is the instantaneous radius of the shell. When the computation has been stopped, the column density along the $\theta=\pi$ lies within the range of $N_{\rm sh}\sim n_{\rm sh}\Delta R_{\rm s}\sim 2\times 10^{23}\,{\rm cm^{-2}}$, where $n_{\rm sh}$ is the number density in the shell. Note that this column density does not increase as expected in the snowplow phase of expansion. The reason for this is that the shocked material in the shell flows back towards the disk midplane, as was sketched in Figure\,\ref{fig:sketch}.
 
We compare this column density with the column density of the remaining DM halo gas outside the shell, $N_{\rm h}$, beyond $R\sim 1$\,pc. Thus estimating the column density of the halo from $R_{\rm sh}\sim 1$\,pc to its virial radius, $R_{\rm h}\sim 600$\,pc, although the outer limit is not important because of the rapid decrease in the ambient gas density. The density profile in the halo is $n_{\rm h0}\sim 6\times 10^4 (R/1\,{\rm pc})^{-2}$, as shown in Figure\,\ref{fig:initial} \citep[in agreement with][]{luo18,ardaneh18}, with the density at $R_{\rm h}$, $n_{\rm h0}\sim 0.1\,{\rm cm^{-3}}$. Integrating over $dN_{\rm h}\sim n_{\rm h}(R) dR$, we obtain $N_{\rm h} \sim 2\times 10^{23}\,{\rm cm^{-2}}$. This is comparable to the column density of the shell.

Next, we calculate the linear momentum per unit surface area of the expanding shell at $t=3.2\times 10^3$\,yr, which is $P_{\rm sh}\sim m_{\rm p} N_{\rm sh} v_{\rm sh}\sim 7\times 10^6\,{\rm g\,cm^{-1}\,s^{-1}}$. At the same time, the linear momentum of the DM halo accretion flow outside the shell is $P_{\rm h}\sim m_{\rm p} N_{\rm h} v_{\rm h}\sim 3\times 10^5\,{\rm g\,cm^{-1}\,s^{-1}}$. Hence, the expanding shell has a substantially larger momentum per unit surface area.

As a next step, we ask the question of whether the wind from the SMS and the surrounding accretion disk is momentum or energy driven. For this purpose, we compare the $\dot M_{\rm w} v_{\rm w}$ with $L_{\rm sink}/c$. The latter one is $2-3\times 10^{43}\,{\rm erg\,s^{-1}}/3\times 10^{10}\,{\rm cm\,s^{-1}}\sim 8\times 10^{32}$\,dyne, while the former is $\sim 2\times 10^{32}$\,dyne, using $\dot M\sim 0.06\,{\rm M_\odot\,yr^{-1}}$ at $R\sim 1$\,pc. This means that ($\dot M_{\rm w} v_{\rm w})/(L_{\rm sink}/c) \sim 0.5$, i.e., the shocked wind can be driven by the linear momentum of radiation output in the system with a non-negligible contribution from the thermal gradients, and the outflow roughly conserves the linear momentum.

Finally, we calculate the kinetic energies per unit surface area of the expanding shell and that of the accreting material outside the shell, at this time. The kinetic energy per unit area of the shell is $E_{\rm kin,sh} \sim \Sigma_{\rm sh} v_{\rm sh}^2 \sim 10^{14}\,{\rm g\,s^{-2}}$. For the halo material, it is $E_{\rm kin,h} \sim \Sigma_{\rm h} v_{\rm h}^2 \sim 3\times 10^{11}\,{\rm g\,s^{-2}}$. Hence kinetic energy per unit surface of the expanding shell is substantially higher than that of the accretion flow.

In summary, while the column densities of the outer shell and the remaining halo gas outside the inner parsec are comparable, the linear momentum and the kinetic energy of the shell per unit area dominate that of the outer halo gas. Given that the energy can be radiated away, the prevalence of linear momentum means that the shell is momentum-driven. Note also that the hot gas within the cavity, which acts as a piston, is heated mainly by the fast shocks propagating from the center outwards, thus assuring that the high temperature of the gas will not be reduced.

The above estimates of the extrapolated kinematics of the wind driven by radiation, including the photoionization energy input, lead us to conclude that the expanding shell will have an effect well beyond the central parsec, and probably within the entire parent DM halo. The characteristic timescale for this breakout is $t\ltorder 10^7$\,yr. The funnel which formed initially along the angular momentum axis is further evacuated by the hydrodynamical collimated wind from the central mass accumulation, the SMS or self-gravitating disk. 

\citet{botella22} have investigated the BH feedback in the super-critical regime, including the radiation force. While not directly comparable to our results due to the various assumption used in their simulation, the outflow on a scale of $\sim 1$\,pc behaves similarly to our extrapolated flow. The outflow velocity becomes flat and is expected to have an impact on the gas within the parent DM halo.

Why does our conclusion differ from that of \citet{regan19}? Again, a direct comparison between these two models is difficult, but a few main differences can be pointed out. First, \citet{regan19} have resorted only to mechanical feedback. Second, they used feedback from a subgrid accretion disk around the SMBH seed, assuming the radiative efficiency of 0.1. We argue that for the SMS or self-gravitating accretion disk, the efficiency is higher, $\eta\sim 1$, because the photons are not absorbed by the horizon of the BH. Third, and most importantly, they did not incorporate the feedback by the ionizing radiation of $\gtorder 13.6$\,eV, further limiting it to being optically thin (e.g., see their section\,2.4.1). Lastly, the parameters of the jet used by them are imposed, while we do not incorporate any subgrid physics in the outflow. 

On the other hand, our simulation has been terminated after the outflow has reached $\sim 1$\,pc, because it is time consuming. Therefore, we only can extrapolate the properties of the interaction between the outflow and accretion flow outside this region. This extrapolation relies on the comparison of the surface densities, linear momenta, and kinetic energies of the outflow and the properties of the accretion flow in the outer halo. It also accounts for the backflow along the funnel that we have observed --- this backflow limits the growth of the surface density of the outer shell in the funnel, thus precluding its fragmentation, i.e., Rayleigh-Taylor or Jeans instabilities. 

\section{Concluding Remarks}
\label{sec:conclusion}

We have modeled the pre-SMBH seed formation processes during direct collapse within DM halos at high redshifts using high-resolution zoom-in cosmological simulations. The precursors of the SMBH seeds have been discussed in the literature and can invoke supermassive stars (SMS) or self-gravitating disks. Accretion on these objects will result in mildly super-critical luminosity which is capable of driving powerful outflows. Radiation transfer and thermal gradients have been calculated using the ray-tracing algorithm. We followed the evolution of these outflows and their interaction with the accretion flow. A number of conclusions can be drawn.

Conditions associated with direct collapse appear unique in some way --- the accretion rate is determined by the DM halo potential, and in the early stages of central mass accumulation will be mildly super-critical. As long as the parent halo can maintain this accretion rate and the central mass does not compete with the star formation, the central object will produce mildly super-critical luminosity. Before the formation of the SMBH seed, this central mass, either in the form of an accretion disk around the SMS or a self-gravitating disk, will generate accretion luminosity which peaks in the UV and some fraction of radiation will also contribute to the soft X-rays. Also, in the case of the SMS, additional luminosity at the Eddington limit is expected from thermonuclear burning. 

We find that the radiation and thermal feedback have a modest effect on the accretion flow causing high-frequency jitter in the accretion rate. However, it does not terminate it or significantly reduce it. On the other hand, the presence of angular momentum in the accretion flow in tandem with the feedback has led to the formation of hot cavities expanding along the disk rotation axis and sideways, resulting in the twin funnels --- the outflow is collimated by the accretion. The funnels are being emptied by fast $\sim 10^3\,{\rm km\,s^{-1}}$ winds which drive dense shells. These winds appear to be mass-loaded from the ablation of the accretion flow. The outermost shells expand along the funnels and sideways, and maintain velocities well above the escape velocity at $\sim 1$\,pc. The column densities of these shells, $\sim 10^{23}\,{\rm cm^{-2}}$, do not evolve during the simulation --- they avoid the snowplow phase because the shocked material flows backward. Finally, the sideway expansion of these cavities can induce additional larger amplitude variability in the accretion flow.

The formation of these funnels has observational corollaries. Firstly, they raise the possibility that these outflows can have an effect on the scales of the parent DM halos. Secondly, extrapolation of the shell kinematics, based on a comparison of its column density to that of the remaining DM halo gas, points to the feasibility of the breakout of this outflow from parsec scales and maybe from the DM halo as well, thus producing a plausible escape route for a subsequent radiation and mechanical feedback. Third, the anisotropic density distribution resulting from these processes can have a strong effect on the detection of Ly$\alpha$ radiation from these objects and can assure the survival of H$_2$ in the accretion flow --- this will be addressed elsewhere. 
 
\section*{Acknowledgements}
We thank the Enzo and YT support team for help. All the analysis has been conducted using yt \citep{turk11}, http://yt-project.org/. I.S. is grateful to Mitch Begelman for discussions on various topics relevant to this work. Y.L. acknowledges the support from the NSFC grants No. 11903026, 12273031. I.S. acknowledges the hospitality of KITP at UCSB where part of this research has been conducted under the NSF under Grant No. NSF PHY-1748958.  This work has been partially supported by the Hubble Theory grant HST-AR-18584 and by JSPS KAKENHI grant 16H02163 (to I.S.) and 19H05810, 20H0018 (to K.N.). I.S. and K.N. are grateful for generous support from the International Joint Research Promotion Program at Osaka University. The STScI is operated by the AURA, Inc., under NASA contract NAS5-26555. The simulations were carried out TianHe-1A at the National Supercomputer Center in Tianjin. 




\bibliographystyle{mnras}
\input{ms.bbl}







\label{lastpage}
\end{document}